%% file: FSQ-12-005_temp.tex
\pdfoutput=1

\documentclass[11pt,twoside,a4paper,cmspaper,final,collab]{cms-tdr}
\def\svnVersion{296409}\def\cmsCernNo{2013-037}\def\cmsCernDate{\today}\def\cmsMessage{Published in Physical Review D as \href{http://dx.doi.org/10.1103/PhysRevD.92.012003}{\doi{10.1103/PhysRevD.92.012003.}}}

\begin{document}\cmsNoteHeader{FSQ-12-005}

\hyphenation{had-ron-i-za-tion}
\hyphenation{cal-or-i-me-ter}
\hyphenation{de-vices}
\RCS$Revision: 293610 $
\RCS$HeadURL: svn+ssh://svn.cern.ch/reps/tdr2/papers/FSQ-12-005/trunk/FSQ-12-005.tex $
\RCS$Id: FSQ-12-005.tex 293610 2015-06-22 12:46:16Z robertc $
\newlength\cmsFigWidth
\ifthenelse{\boolean{cms@external}}{\setlength\cmsFigWidth{0.49\textwidth}}{\setlength\cmsFigWidth{0.75\textwidth}}
\newlength\cmsFigWidthN
\ifthenelse{\boolean{cms@external}}{\setlength\cmsFigWidthN{0.49\textwidth}}{\setlength\cmsFigWidthN{0.65\textwidth}}
\ifthenelse{\boolean{cms@external}}{\providecommand{\cmsLeft}{top}}{\providecommand{\cmsLeft}{left}}
\ifthenelse{\boolean{cms@external}}{\providecommand{\cmsRight}{bottom}}{\providecommand{\cmsRight}{right}}
\providecommand{\cPX}{\ensuremath{X}\xspace}
\providecommand{\cPY}{\ensuremath{Y}\xspace}
\providecommand{\PYTHIAMBR}{\PYTHIA8~MBR\xspace}
\providecommand{\QGSJETII}{\textsc{qgsjet-ii}\xspace}
\providecommand{\EPOS}{\textsc{epos}\xspace}
\providecommand{\PHOJET}{\textsc{phojet}\xspace}
\ifthenelse{\boolean{cms@external}}{\providecommand{\NA}{\ensuremath{\cdots}\xspace}}{\providecommand{\NA}{\ensuremath{\text{---}}\xspace}}
\cmsNoteHeader{FSQ-12-005}
\title{Measurement of diffractive dissociation cross sections in \texorpdfstring{$\Pp\Pp$}{pp} collisions at \texorpdfstring{$\sqrt{s}=7$\TeV}{sqrt(s)=7 TeV}}

\date{\today}

\abstract{
Measurements of diffractive dissociation cross sections in $\Pp\Pp$ collisions at $\sqrt{s}=7$\TeV are presented in kinematic regions defined by the masses $M_{\cPX}$ and $M_{\cPY}$ of the two final-state hadronic systems separated by the largest rapidity gap in the event.  Differential cross sections are measured as a function of $\xi_{\cPX}= M^2_{\cPX}/s$ in the region $-5.5<\log_{10}\xi_{\cPX}<-2.5$, for $\log_{10}M_{\cPY}<0.5$, dominated by single dissociation (SD), and $0.5<\log_{10}M_{\cPY}<1.1$, dominated by double dissociation (DD), where $M_{\cPX}$ and $M_{\cPY}$ are given in \GeVns. The inclusive $\Pp\Pp$ cross section is also measured as a function of the width of the central pseudorapidity gap $\Delta\eta$ for $\Delta\eta>3$, $\log_{10}M_{\cPX}>1.1$, and $\log_{10}M_{\cPY}>1.1$, a region dominated by DD. The cross sections integrated over these regions are found to be, respectively, $2.99 \pm 0.02\stat {}_{-0.29}^{+0.32}\syst$\unit{mb}, $1.18 \pm 0.02\stat\pm 0.13\syst$\unit{mb}, and $0.58 \pm 0.01\stat {}_{-0.11}^{+0.13}\syst$\unit{mb}, and are used to extract extrapolated total SD and DD cross sections. In addition, the inclusive differential cross section, $\rd\sigma / \rd\Delta\eta^\mathrm{F}$, for events with a pseudorapidity gap adjacent to the edge of the detector, is measured over $\Delta\eta^\mathrm{F}$ = 8.4 units of pseudorapidity. The results are compared to those of other experiments and to theoretical predictions, and found compatible with slowly-rising diffractive cross sections as a function of center-of-mass energy.
}

\hypersetup{%
pdfauthor={CMS Collaboration},%
pdftitle={Measurement of diffraction dissociation cross sections in pp collisions at sqrt(s)=7 TeV},%
pdfsubject={CMS},%
pdfkeywords={CMS, physics, diffraction, rapidity gaps}}

\maketitle

\section{Introduction}
\label{sec:introduction}
\begin{figure*}[bht]
\centering
\includegraphics[width=\textwidth]{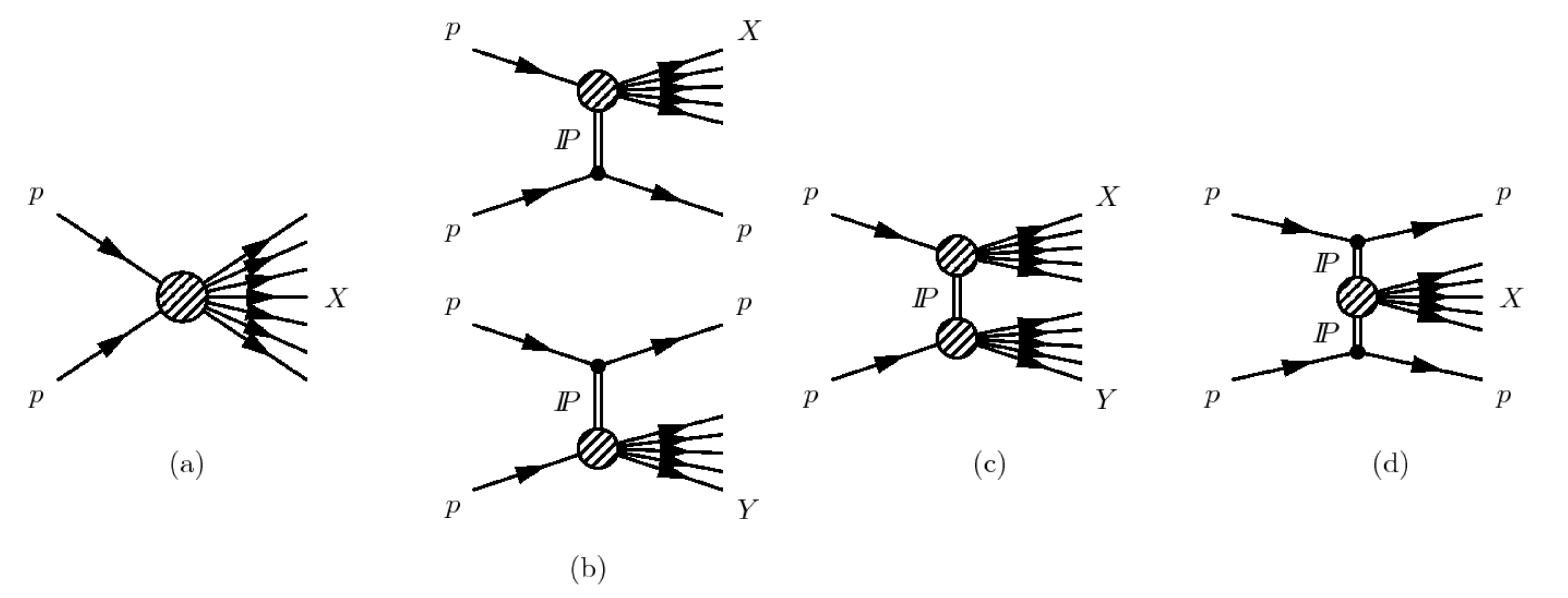}
\caption{Schematic diagrams of (a) nondiffractive, $\Pp\Pp\to \cPX$, and diffractive processes with (b) single dissociation, $\Pp\Pp\to \cPX\Pp$ or $\Pp\Pp\to \Pp \cPY$, (c) double dissociation, $\Pp\Pp\to \cPX\cPY$, and (d) central diffraction, $\Pp\Pp\to \Pp \cPX\Pp$; $\cPX$($\cPY$) represents a dissociated proton or a centrally produced hadronic system.}
\label{fig:1}
\end{figure*}

A significant fraction ($\approx$25\%) of the total inelastic proton--proton cross section at high energies can be attributed to diffractive interactions, characterized by the presence of at least one non-exponentially suppressed large rapidity gap (LRG), i.e. a region of pseudorapidity $\eta$ devoid of particles, where for a particle moving at a polar angle $\theta$ with respect to the beam $\eta=-\ln[\tan (\theta/2)]$. If this $\eta$ region is adjacent to the diffractively scattered proton it is called a forward pseudorapidity gap. In hadronic interactions an LRG is presumed to be mediated by a color-singlet exchange carrying the vacuum quantum numbers, commonly referred to as Pomeron exchange. Figure~\ref{fig:1} defines the main types of diffractive processes: single dissociation (SD), double dissociation (DD), and central diffraction (CD).

Inclusive diffractive cross sections cannot be calculated within perturbative quantum chromodynamics, and are commonly described by models based on Regge theory (see \eg~\cite{Donnachie:2002en} and references therein). The predictions of these models generally differ when extrapolated from the Tevatron center-of-mass energies of $\sqrt{s} \le 1.96$\TeV to LHC energies. Therefore, measurements of diffractive cross sections at 7\TeV provide a valuable input for understanding diffraction and improving its theoretical description. They are also crucial for the proper modeling of the full final state of hadronic interactions in event generators, and can help to improve the simulation of the underlying event, as well as of the total inelastic cross section.

The DD cross section has been recently measured at $\sqrt{s}=7$\TeV by the TOTEM collaboration~\cite{Antchev:2013any}, for events in which both dissociated-proton masses are below $\sim$12\GeV. Other measurements of diffractive cross sections at the LHC, with higher dissociation masses, have either a limited precision~\cite{Abelev:2012sea} or no separation between SD and DD events~\cite{Aad:2012pw}. In this paper, we present the first CMS measurement of inclusive diffractive cross sections at $\sqrt{s}=7$\TeV. This measurement is based on the presence of a forward LRG, with SD- and DD-dominated event samples separated by using the CASTOR calorimeter~\cite{Andreev:2010zzb}, covering the very forward region, $-6.6 < \eta < -5.2$. A data sample with a central LRG, in which DD dominates, is also used. In addition, the inclusive differential cross section, $\rd\sigma / \rd\Delta\eta^\mathrm{F}$, for events with a pseudorapidity gap adjacent to the edge of the detector, is measured over $\Delta\eta^\mathrm{F}$ = 8.4 units of pseudorapidity, and compared to a similar ATLAS measurement~\cite{Aad:2012pw}. The results presented here are based on the first CMS data collected at $\sqrt{s}=7$\TeV during the 2010 LHC commissioning period, when the probability of overlapping $\Pp\Pp$ interactions in the same bunch crossing (pileup), which may spoil the detection of the gap, was low.

The paper is organized in eleven sections and two appendices. The CMS detector is described in Section~\ref{sec:det}. Section~\ref{sec:mc} presents the Monte Carlo (MC) simulations used in the analysis. The event selection and the diffractive event topologies used to measure the cross sections are discussed in Sections~\ref{sec:selection} and \ref{sec:topo}, respectively. Sections~\ref{sec:sd}, \ref{sec:dd}, and \ref{sec:integsig} present the measurement of the forward-gap and central-gap differential cross sections, and the integrated cross sections, respectively. All cross sections are extracted within the detector acceptance and with minimal model-dependent systematic uncertainties. The extrapolation of the measured cross sections to the low mass regions is discussed in Section~\ref{sec:sdddsig}. Section~\ref{sec:rapgapxsec} presents the  measurement of the pseudorapidity gap cross section and its comparison to the ATLAS result~\cite{Aad:2012pw}. The systematic uncertainties for all the measurements are discussed in Section~\ref{sec:syst}. A summary is given in Section~\ref{sec:sum}. Appendices A and B present additional comparisons between the diffractive MC models used.

\section{The CMS detector}
\label{sec:det}

A detailed description of the CMS detector, together with a definition of the coordinate system used and the relevant kinematic variables, can be found in Ref.~\cite{Chatrchyan:2008zzk}. The central feature of the apparatus is a superconducting solenoid of 6 m internal diameter, providing a 3.8\unit{T} axial field. Within the field volume are located a silicon pixel and strip tracker, a crystal electromagnetic calorimeter (ECAL), and a brass/scintillator hadron calorimeter. Muons are measured in gas-ionization detectors embedded in the steel flux-return yoke of the magnet. The calorimeter cells are grouped in projective towers, of granularity $\Delta\eta \times \Delta \phi = 0.087 \times 0.087$ at central rapidities and $0.175 \times 0.175$ at forward rapidities. In addition to the barrel and endcap detectors, CMS has extensive forward calorimetry. The forward component of the hadron calorimeter, HF ($2.9 < \abs{\eta} < 5.2$), consists of steel absorbers with embedded radiation-hard quartz fibers, providing fast collection of Cherenkov light. The very forward angles are covered at one end of CMS ($-6.6 < \eta < -5.2$) by the CASTOR calorimeter~\cite{Andreev:2010zzb}, made of quartz plates embedded in tungsten absorbers, segmented in 16 $\phi$-sectors and 14 $z$-modules.

Two elements of the CMS monitoring system, the Beam Scintillator Counters (BSC) and the Beam Pick-up Timing eXperiment (BPTX) devices, are used to trigger the CMS readout. The two BSC are located at a distance of $\pm$10.86\unit{m} from the nominal interaction point (IP) and are sensitive in the $\abs{\eta}$ range 3.23 to 4.65. Each BSC consists of 16 scintillator tiles. The BSC elements have a time resolution of 3\unit{ns} and an average minimum ionizing particle detection efficiency of 96.3\%. The two BPTX devices, located around the beam pipe at a distance of 175\unit{m} from the IP on either side, are designed to provide precise information on the bunch structure and timing of the incoming beams, with better than 0.2 ns time resolution.

\section{Monte Carlo simulation}
\label{sec:mc}

Monte Carlo simulations are used to correct the measured distributions for the geometrical acceptance and reconstruction efficiency of the CMS detector, as well as for migrations from true to reconstructed values in the distributions of kinematic variables. We use \PYTHIA8.165~\cite{Sjostrand:2006za,Sjostrand:2007gs} to generate samples of inelastic events. We compare the detector-level data distributions to the \PYTHIA8~4C~\cite{Sjostrand:2007gs} and \PYTHIAMBR (Minimum Bias Rockefeller)~\cite{mbrnote} simulations and extract integrated cross sections using \PYTHIAMBR.

Diffractive events in the \PYTHIA8~4C simulation are generated according to the Schuler--Sj\"ostrand model implemented in \PYTHIA6~\cite{Sjostrand:2006za}. The 4C tune~\cite{Sjostrand:2007gs} includes a downward scaling of the Schuler--Sj\"ostrand SD and DD cross sections at $\sqrt{s}= 7$\TeV by 10 and 12\%, respectively.

The \PYTHIAMBR generator predicts the energy dependence of the total, elastic, and inelastic $\Pp\Pp$ cross sections, and fully simulates the main diffractive components of the inelastic cross section: SD, DD and CD. The diffractive-event generation in \PYTHIAMBR is based on a phenomenological renormalized--Regge--theory model~\cite{dinoModel,dinoModel2}, which is unitarized by interpreting the Pomeron flux as the probability for forming a diffractive rapidity gap. The model was originally developed for the CDF experiment at the Tevatron and has been successfully tested with the CDF results on diffraction. The \PYTHIAMBR simulation assumes a linear parametrization of the Pomeron trajectory, $\alpha(t)=1+\varepsilon+\alpha't$, where $t$ is the square of the four-momentum transfer between the two incident protons. We use $\alpha'= 0.25\GeV^{-2}$, and $\varepsilon=0.08$~\cite{Donnachie:1984xq,Donnachie:1992ny} or $\varepsilon=0.104$~\cite{Covolan:1996uy,dinointercept}, to account for the possible energy dependence of $\varepsilon$ in the range of diffractive masses accessible in this analysis. We find that the simulation with $\varepsilon=0.08$ gives a good description of the data. Scaling the DD cross section downwards by 15\%, which preserves the agreement with the CDF data, further improves the description of the DD-dominated data at $\sqrt{s} = 7$\TeV. These modifications are incorporated into the simulation used here.
The measured cross sections are also compared to the predictions of \PYTHIA6~Z2*~\cite{Chatrchyan:2013gfi} and to MC generators based on Regge--Gribov phenomenology:  \PHOJET (version 1.12-35)~\cite{PhojetR,:PHOJET}, \QGSJETII (versions 03 and 04)~\cite{Ostapchenko:2004ss,Ostapchenko:2010vb}, and \EPOS~LHC~\cite{Werner:2005jf}; the latter two are commonly used in cosmic-ray physics~\cite{d'Enterria:2011kwR}.

At the stable-particle level (where stable particles are those with lifetime $\tau$ such that $c\tau >10$\unit{mm}), the kinematic regions covered by the present measurements are defined by the masses $M_\cPX$ and $M_\cPY$ of the two final-state hadronic systems separated by the largest rapidity gap in the event. For a final-state particle of energy $E$ and longitudinal momentum $p_z$, rapidity is defined as $y=(1/2)\ln[(E+p_z)/(E-p_z)]$. At stable-particle level the gap is defined as the largest rapidity separation between stable particles, without any acceptance restriction. The final state is then separated into systems $\cPX$ and $\cPY$, which populate the regions on the positive and negative side of the rapidity gap, respectively. The corresponding masses $M_\cPX$ and $M_\cPY$ are calculated from the full set of four-vectors in the respective group of stable particles. In the following Sections, $M_\cPX$ and $M_\cPY$ are given in units of \GeVns.

We use the pseudorapidity variable to select diffractive events at the detector level.  At the stable-particle level, the true rapidity is used. For the pseudorapidity gap cross section, pseudorapidity (not true rapidity) is used at the hadron level to avoid unnecessary large bin migrations between the distributions measured at the detector and stable-particle levels.  As the central CMS detector is insensitive to low-mass diffraction, we use the \PYTHIAMBR simulation, which describes the data well, to extrapolate the measured cross section into the low mass region.

The detailed MC simulation of the CMS detector response is based on \GEANTfour~\cite{Agostinelli:2002hh}. Simulated  \PYTHIA8~4C and \PYTHIAMBR events are processed and reconstructed in the same manner as collision data.

\section{Event selection}
\label{sec:selection}

The present analysis is based on event samples collected during the 2010 commissioning period, when the LHC was operating at low pileup. For the results presented in Sections~\ref{sec:topo}--\ref{sec:sdddsig}, only data with information from the CASTOR calorimeter are used, which correspond to an integrated luminosity of 16.2\mubinv, and have an average number of inelastic $\Pp\Pp$ collisions per bunch crossing of $\mu=0.14$. The results based on pseudorapidity-gap events presented in Section~\ref{sec:rapgapxsec} are extracted from a different set of data taking runs with negligible pileup ($\mu=0.006$) that correspond to an integrated luminosity of 20.3\mubinv. Events were selected online by requiring a signal in both BPTX detectors, in conjunction with a signal in any of the BSC scintillators. These conditions correspond to requiring the presence of two crossing bunches along with activity in the main CMS detector (minimum bias trigger).

\begin{figure*}[t!hb]
\centering
\includegraphics[width=\textwidth]{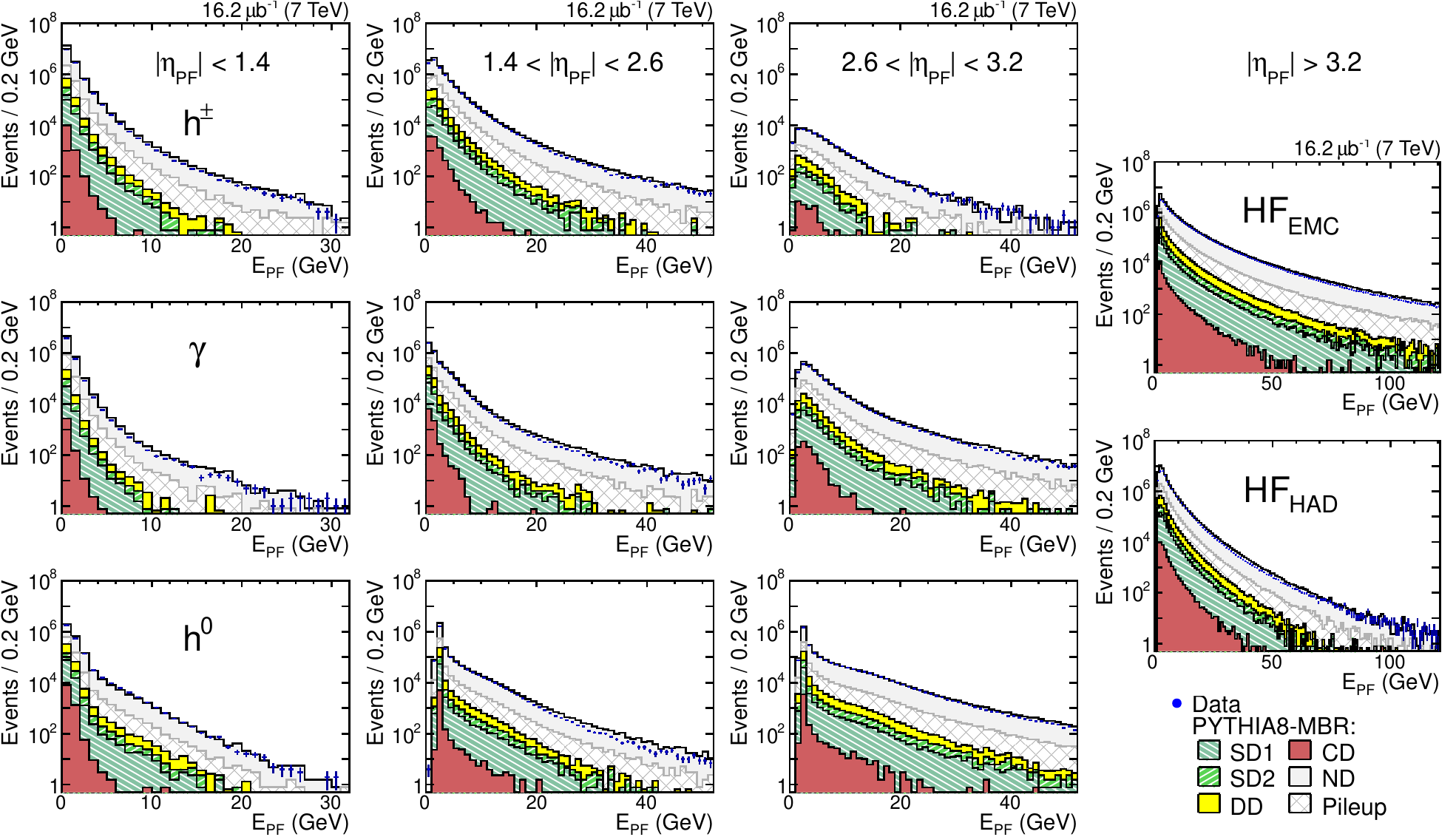}
\caption{Detector-level distributions of the energy of PF objects in four pseudorapidity intervals: $|\eta_{PF}|<1.4$, $1.4<|\eta_{PF}|<2.6$, $2.6<|\eta_{PF}|<3.2$, and $|\eta_{PF}|>3.2$, corresponding to the barrel, endcap, endcap-forward transition, and forward detector regions (columns), for five particle candidate types: charged hadrons (tracks), photons, neutral hadrons, and two types that yield electromagnetic or hadronic energy deposits in HF (rows). Electron and muon candidates constitute less than 0.1\% of the PF objects reconstructed in the $|\eta_{PF}|<2.6$ region, and are not shown. The data are compared to the predictions of the \PYTHIAMBR simulation, normalized to the integrated luminosity of the data sample. The contribution of each of the generated processes is shown separately.}
\label{fig:20}
\end{figure*}

Offline selections \cite{Khachatryan:2010xs} are applied to remove beam-scraping, beam-halo, and noise events. In addition, a minimal activity in the main CMS detectors is imposed offline by requiring at least two particle-flow (PF) objects reconstructed within the geometrical acceptance of the BSC detectors ($3.23<\abs{\eta}<4.65$, with an energy of at least 4\GeV for each PF object). Particle-flow objects~\cite{PF,PF2} are particle candidates obtained by optimally combining the information from the tracking system and the calorimeters. In the forward regions ($\abs{\eta}>2.5$), where there is no tracking, PF objects reduce to calorimeter towers. Figure~\ref{fig:20} shows the distributions of the energy of the PF objects reconstructed in different detector regions for different particle candidates, compared to the prediction of the \PYTHIAMBR simulation, which describes the data well. The requirement on the minimum energy of PF objects was found by studying data collected in dedicated runs with no beam; it depends on the detector region and the particle candidates, and varies from zero for tracks to 4\GeV for the HF towers. To assure a reliable Monte Carlo description of the data, the two innermost (most forward) rings of HF are not used in the analysis, thus limiting the central CMS detector coverage to $\abs{\eta} \lesssim 4.7$. The two outermost (most central) HF rings are also not used for the same reason. No vertex requirement is imposed. This procedure gives high acceptance for diffractive events with the hadronic system outside the tracking acceptance (i.e. with low to moderate diffractive masses, $12 \lesssim M_\cPX \lesssim 100$\GeV). According to the \PYTHIAMBR simulation, the selection described above accepts about 90\% of the events corresponding to the total inelastic cross section in the region of $\log_{10}M_\cPX>1.1$ or $\log_{10}M_\cPY>1.1$.

\section{Diffractive event topologies}
\label{sec:topo}

\begin{figure}[t!b]
\centering
\includegraphics[width=\cmsFigWidth]{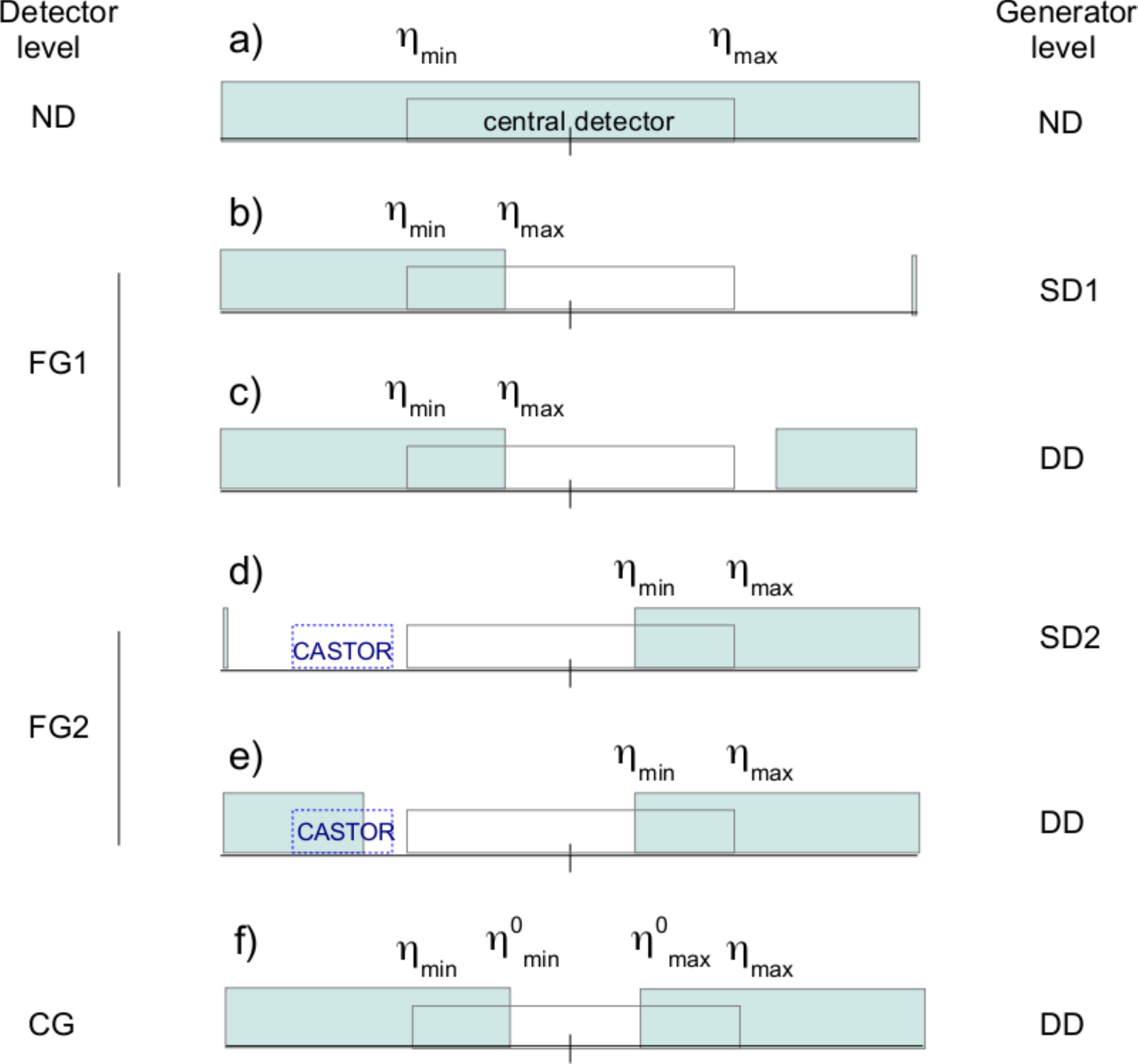}
\caption{Event topologies in final-state particle $\eta$ space. Detector level: nondiffractive events (ND), diffractive events with a forward pseudorapidity gap on the positive (FG1) or negative (FG2) $\eta$-side of the detector, or with a central pseudorapidity gap (CG). Generator level: (a) ND, $\Pp\Pp\to \cPX$, (b) SD1, $\Pp\Pp\to \cPX\Pp$, (d) SD2, $\Pp\Pp\to \Pp \cPY$, and (c, e ,f) DD, $\Pp\Pp\to \mathrm{XY}$, events. The empty box represents the central CMS detector ($\abs{\eta}\lesssim 4.7$), filled full boxes indicate final-state hadronic systems or a proton---the vertical thin bar at the right/left end of sketch (b)/(d). The dotted empty boxes in (d) and (e) represent the CASTOR calorimeter ($-6.6<\eta<-5.2$).}
\label{fig:2}
\end{figure}

The events satisfying the selection described in Section~\ref{sec:selection} constitute a minimum bias sample dominated by inclusive inelastic events in the region covered by the central CMS detector ($\abs{\eta} \lesssim 4.7$). They are mostly composed of nondiffractive (ND) events for which final-state particle production occurs in the entire $\eta$ space available, as shown schematically in Fig.~\ref{fig:2}a. In contrast, diffractive events are expected to have an LRG in the final state. Experimentally, the following diffractive topologies are defined, depending on the position of the reconstructed LRG in the central detector:

\begin{figure}[t!hp]
\centering
\includegraphics[width=\cmsFigWidthN]{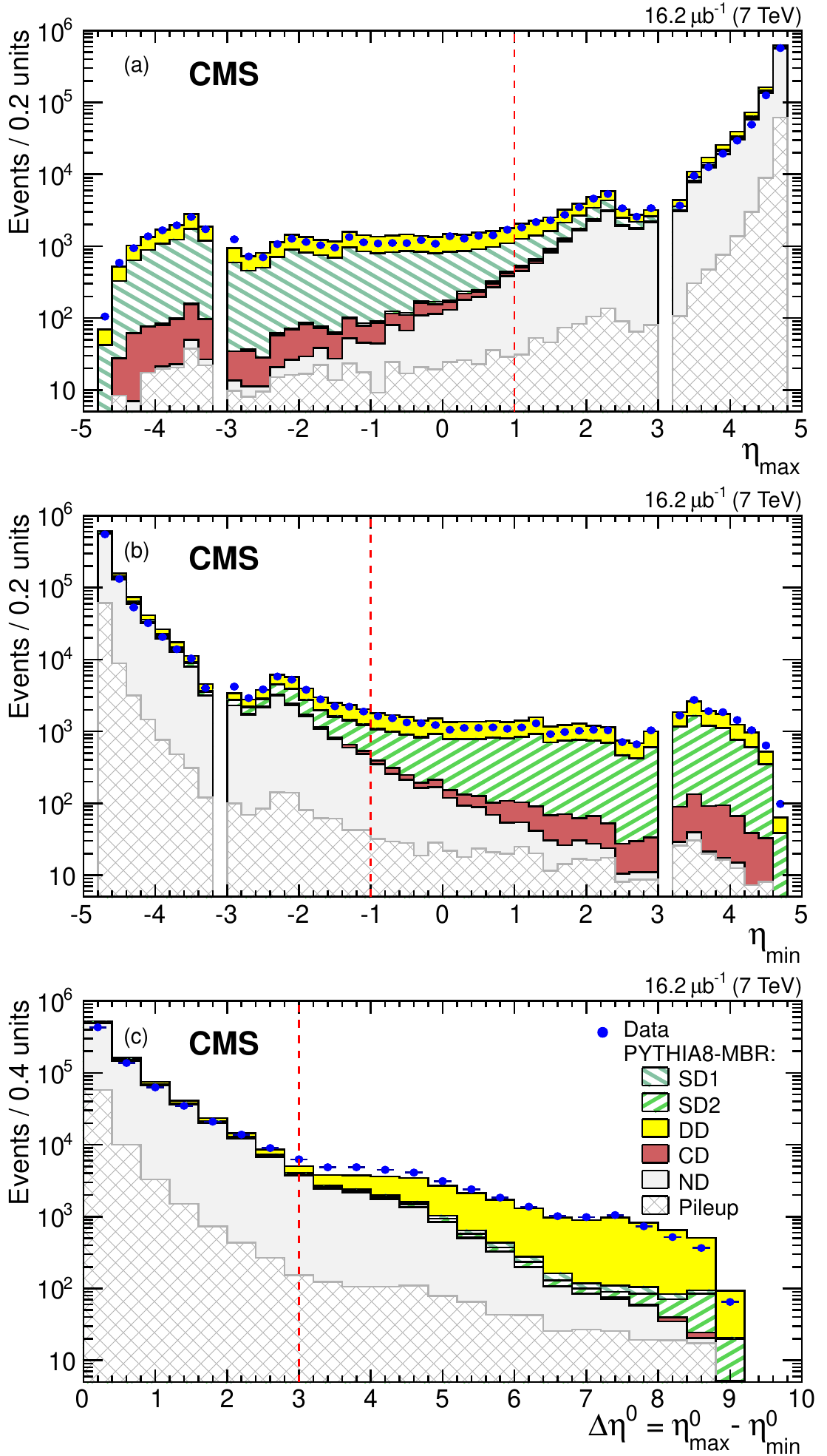}
\caption{Detector-level distributions for the (a) $\eta_\text{max}$, (b) $\eta_\text{min}$, and (c) $\Delta\eta^{0} = \eta^{0}_\text{max}-\eta^{0}_\text{min}$ variables measured in the minimum bias sample (with only statistical errors shown), compared to predictions of the \PYTHIAMBR simulation normalized to the integrated luminosity of the data sample. Contributions from each of the MC-generated processes, and simulated events with at least two overlapping interactions of any type (pileup), are shown separately. The dashed vertical lines indicate the boundaries for the $\eta_\text{max}<1$, $\eta_\text{min}>-1$, and  $\Delta\eta^{0}>3$ selections.}
\label{fig:3}
\end{figure}

\begin{itemize}
\item FG1: a forward pseudorapidity gap at the edge of the detector on the positive $\eta$-side (Figs.~\ref{fig:2}b,c);
\item FG2: a forward pseudorapidity gap at the edge of the detector on the negative $\eta$-side (Figs.~\ref{fig:2}d,e);
\item CG: a central pseudorapidity gap in the detector around $\eta = 0$ (Fig.~\ref{fig:2}f).
\end{itemize}

The experimental topology with a forward pseudorapidity gap on each edge of the detector (CD topology, Fig.~\ref{fig:1}d) is neglected in this analysis because of the limited number of such events. For the FG1 and FG2 topologies the pseudorapidity gap is related to the variables $\eta_\text{max}$ and $\eta_\text{min}$ (Fig.~\ref{fig:2}), defined as the highest (lowest) $\eta$ of the PF object reconstructed in the central detector. Experimentally, the pseudorapidity gap in CG events may be expressed as $\Delta\eta^{0} = \eta^{0}_\text{max}-\eta^{0}_\text{min}$, where $\eta^{0}_\text{max}$ ($\eta^0_\text{min}$) is the closest-to-zero $\eta$ value of the PF objects reconstructed on the positive (negative) $\eta$-side of the central detector (Fig.~\ref{fig:2}).

Figure~\ref{fig:3} shows the distributions of $\eta_\text{max}$, $\eta_\text{min}$, and $\Delta\eta^{0} = \eta^{0}_\text{max}-\eta^{0}_\text{min}$ for the minimum bias sample defined in Section~\ref{sec:selection}, compared to MC predictions. For the $\Delta\eta^{0}$ selection, the additional requirement that activity be present on both $\eta$-sides of the central detector is imposed. The data are dominated by the contribution from ND events, for which rapidity gaps are exponentially suppressed~\cite{PhysRevD.47.101}. Diffractive events appear as a flattening of the exponential distributions, and dominate the regions of low $\eta_\text{max}$, high $\eta_\text{min}$, and high $\Delta\eta^{0}$. The absence of events around $\abs{\eta_\text{max}}(\abs{\eta_\text{min}})\approx3$ in Fig.~\ref{fig:3}a (\ref{fig:3}b) reflects the fact that the two outermost (most central) rings of HF are not used in the analysis; the depletion of events around $\abs{\eta_\text{max}}(\abs{\eta_\text{min}})\approx2.4$ corresponds to the transition region between the tracker and the forward calorimeters, where higher thresholds are applied for the latter. The regions of $3 \lesssim \Delta \eta^{0}\lesssim 6$ and $\Delta \eta^{0}\gtrsim 6$ in Fig.~\ref{fig:3}c correspond to the DD topology for which one or both of the $\eta^{0}_\text{max}$ and $\eta^{0}_\text{min}$ edges are in the HF calorimeters. In order to select samples of FG1, FG2, and CG events with a central LRG signature, the requirements $\eta_\text{max}<1$, $\eta_\text{min}>-1$, and $\Delta \eta^{0}>3$ are imposed, respectively.

According to the expectations of the \PYTHIAMBR simulation, the event samples defined experimentally as FG1 or FG2 (Figs.~\ref{fig:3}a,b) originate from approximately equal numbers of SD events with $1.1 \lesssim \log_{10}M_\cPX \lesssim 2.5$ and DD events for which one dissociated-proton mass is in this $M_\cPX$ range, while the other is small and escapes detection in the central detector, cf. Figs.~\ref{fig:2}c,e. For the FG2 topology, CASTOR ($-6.6 < \eta < -5.2$) is used to separate diffractive events into two samples: $ \log_{10}M_\cPY \lesssim 0.5$ (SD enhanced) and $ 0.5 \lesssim \log_{10}M_\cPY \lesssim 1.1$ (DD enhanced, Figs.~\ref{fig:2}d,e). The detection of the low-mass dissociated system, $Y$, is performed by using a CASTOR tag, defined as the presence of a signal above the energy threshold (1.48\GeV) in at least one of the 16 $\phi$-sectors of the first five CASTOR modules. Since no detector is available for tagging the low-mass dissociated system on the positive $\eta$-side, the FG1 sample is treated as a control sample in this analysis.

\begin{figure}[th]
\centering
\includegraphics[width=0.48\textwidth]{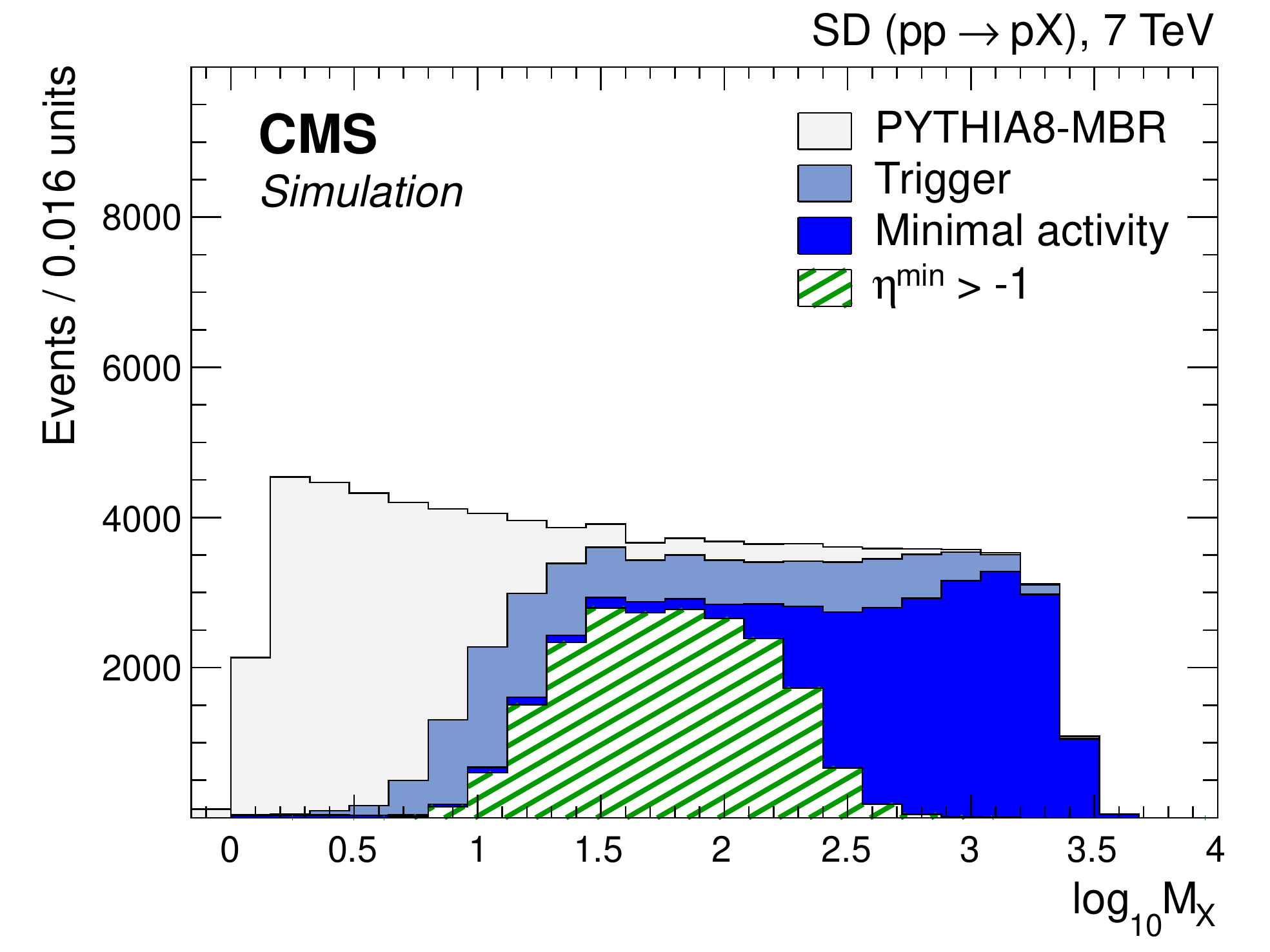}
\includegraphics[width=0.48\textwidth]{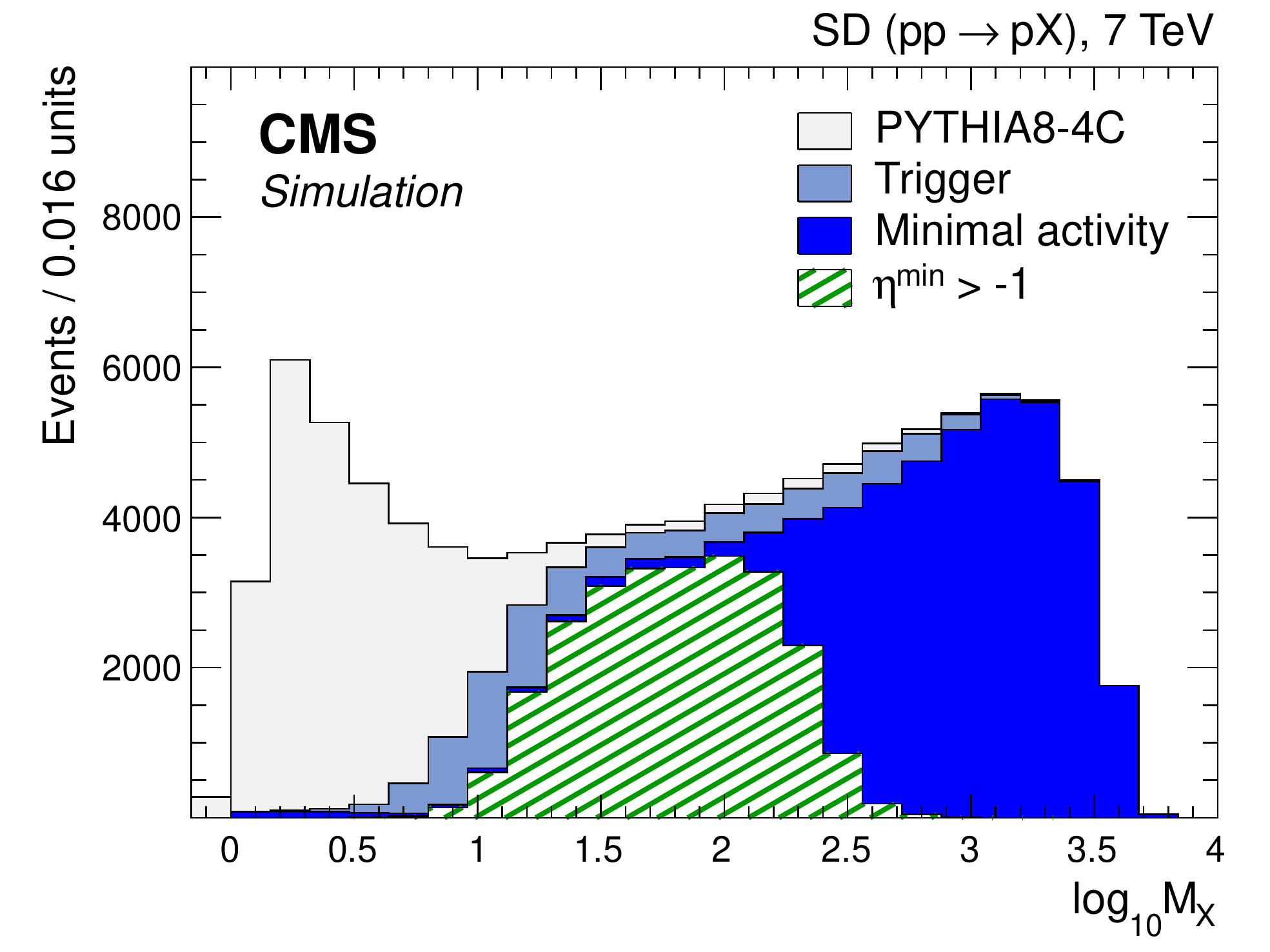}
\caption{Simulated distributions of the dissociated mass $M_\cPX$ at stable-particle level for the SD process in the FG2 sample at successive selection stages (trigger, minimal detector activity within BSC acceptance, $\eta_\text{min}>-1$) for \PYTHIAMBR (\cmsLeft) and \PYTHIA8~4C (\cmsRight). The MC samples are normalized to the luminosity of the data sample.}
\label{fig:4a}
\end{figure}

\begin{figure*}[th]
\centering
\includegraphics[width=\textwidth]{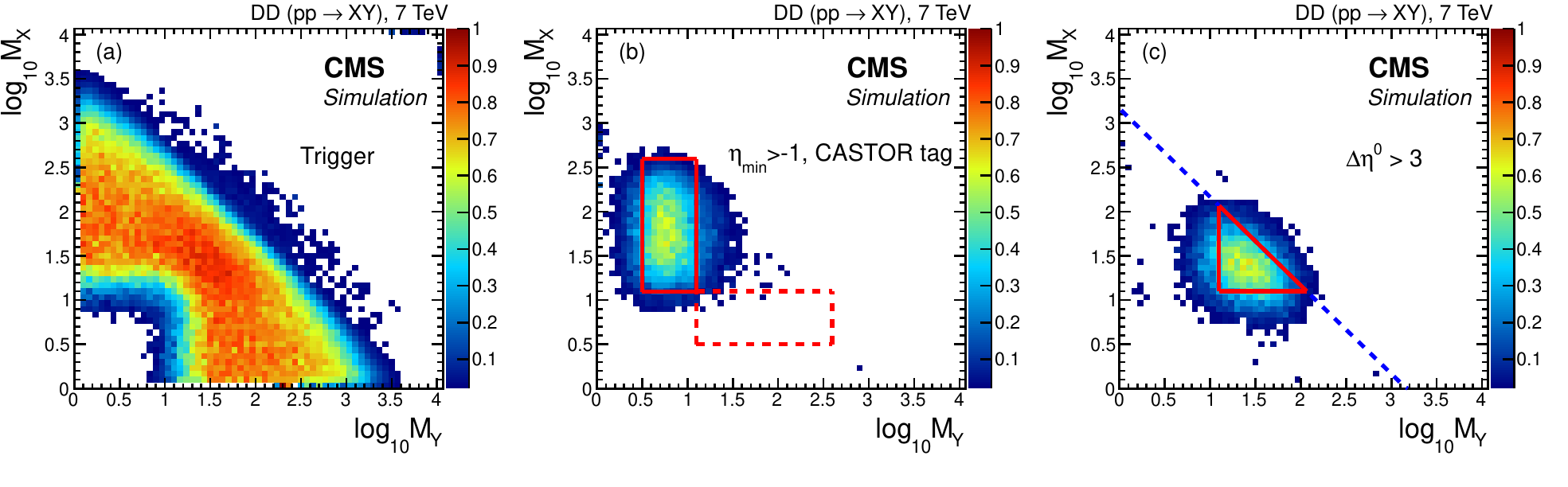}
\caption{Simulated  (\PYTHIAMBR) event selection efficiency in the $M_\cPX$ vs. $M_\cPY$ plane for true DD events after (a) the trigger selection, and (b) the FG2 selection with a CASTOR tag or (c) the CG selection (Fig.~\ref{fig:2}). The regions delimited by the solid (red) lines in (b) and (c) are those of the cross section measurements; the dashed (red) box in (b) corresponds to the enlarged region for which the cross section is given (Section~\ref{sec:sdddsig}), assuming the same dependence on $M_\cPX$ and $M_\cPY$; the dashed (blue) line in (c) marks the region of $\Delta \eta>3$. }
\label{fig:4}
\end{figure*}

The range of the dissociation mass $M_\cPX$ for the true SD process in the FG2-type sample after all detector selections is shown as a hatched histogram in Fig.~\ref{fig:4a} for \PYTHIAMBR (\cmsLeft) and \PYTHIA8~4C (\cmsRight), and corresponds to $ 1.1 \lesssim \log_{10}M_\cPX \lesssim 2.5$. Similar distributions are obtained for events in the FG1-type sample, in which the dissociated system originates from the proton on the other side of the detector. The ranges of dissociation masses, $M_\cPX$ and $M_\cPY$, for the true DD events in the minimum bias sample after the trigger selection, in the FG2-type sample with a CASTOR tag, and in the CG-type sample after all detector selections, are shown in the efficiency plots of Figs.~\ref{fig:4}a, \ref{fig:4}b,  and \ref{fig:4}c, respectively. The FG2-type events, with the pseudorapidity gap reconstructed at the edge of the central detector, populate the region of $1.1 \lesssim \log_{10}M_\cPX \lesssim 2.5$ and $ 0.5 \lesssim \log_{10}M_\cPY \lesssim 1.1$ (solid box in Fig.~\ref{fig:4}b). The selection based on the CG topology requires both diffractive masses to be in the central detector; this leads to different coverage in the ($M_\cPX$, $M_\cPY$) plane. Events populate the region of  $\log_{10}M_\cPX \gtrsim 1.1$ and $\log_{10}M_\cPY \gtrsim 1.1$ (Fig.~\ref{fig:4}c), in addition to $\Delta \eta^{0}>3$, thus providing a complementary measurement of the DD cross section.

\section{Forward pseudorapidity gap cross sections from the FG2 event sample}
\label{sec:sd}

The forward pseudorapidity gap cross sections are measured as a function of the variable $\xi_\cPX$, which is related to the mass $M_\cPX$ of the dissociated system by:
\begin{equation}
\xi_\cPX=\frac{M^2_\cPX}{s}.
\label{eq:xidef}
\end{equation}
For the FG2 sample, $M_\cPX$ corresponds to the dissociated system that can be detected in the central detector (right-hand side of Figs.~\ref{fig:2}d,e). The CASTOR calorimeter allows the detection of the hadronic system $Y$ when it escapes the central detector, and the separation of the FG2 sample into subsamples corresponding to $\log_{10}M_\cPY<0.5$ and $0.5<\log_{10}M_\cPY<1.1$, which are dominated by SD and DD events, respectively. For the purely SD events, $\xi_\cPX$ represents the fractional longitudinal momentum loss of the incoming proton.

At detector level, the variable $\xi_\cPX$ is reconstructed as
\begin{equation}
\xi^{\pm}_\cPX = \frac{\sum_{i}(E^i\mp p^i_z)}{\sqrt{s}},
\label{eq:xirecodef}
\end{equation}
where $i$ runs over all PF objects measured in the central detector, and $E^i$ and $p^i_z$ are the energy and the longitudinal momentum of the $i$th PF object, respectively. The energy is related to the particle three-momentum assuming a mass that depends on the PF object type; \eg for charged hadrons a pion mass is assumed. The signs ($\pm$) in Eq.~(\ref{eq:xirecodef}) indicate whether the dissociated system is on the ${\pm}z$ side of the detector. For the FG2-type events under study, $\xi_\cPX$ corresponds to $\xi^{+}_\cPX$.

\begin{figure}[bht]
\centering
\includegraphics[width=0.49\textwidth]{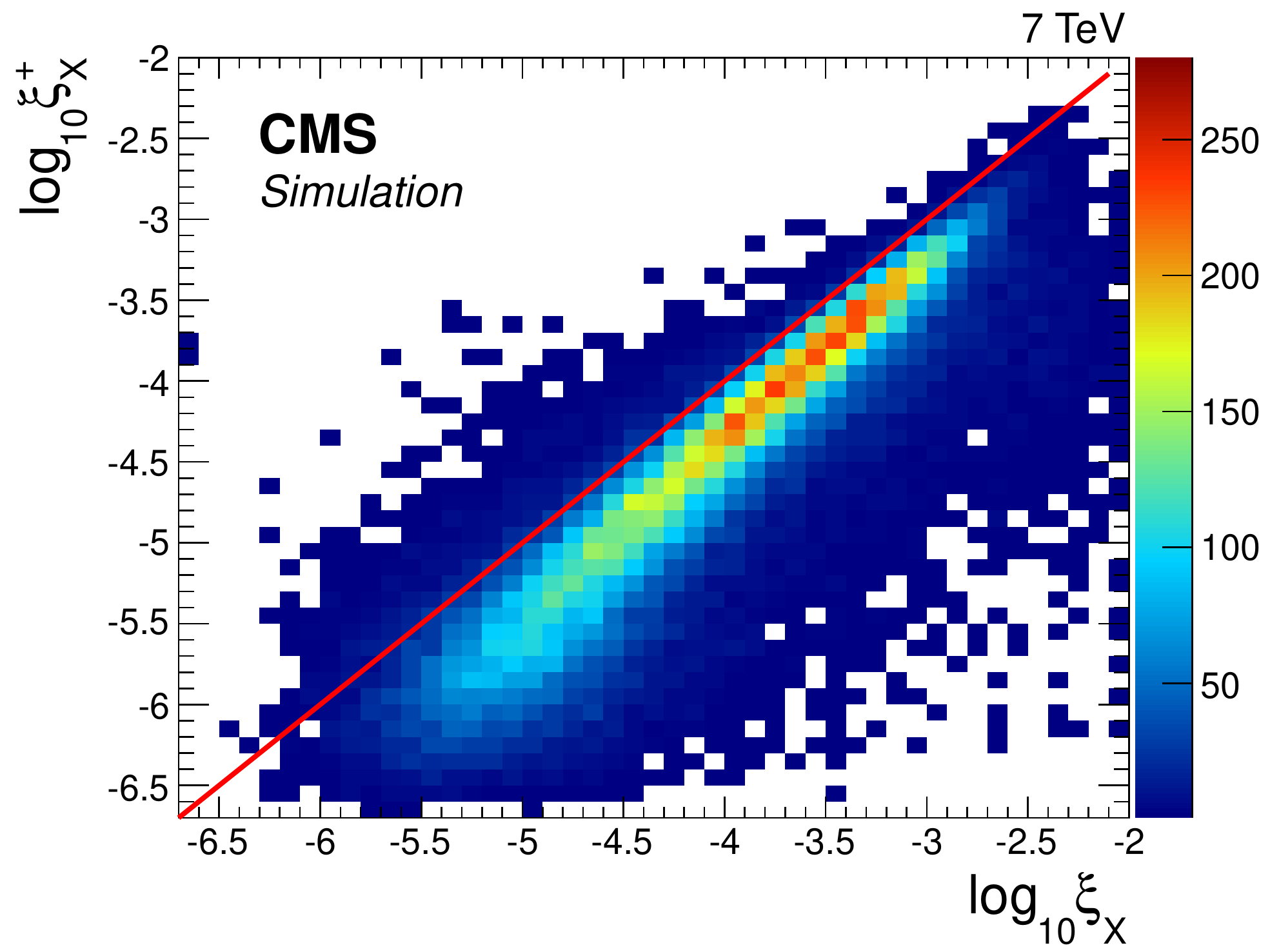}
\caption{Two-dimensional distribution of reconstructed  $\xi^{+}_\cPX$ vs. generated $\xi_\cPX$ values for the events in the SD2 sample obtained with the \PYTHIAMBR simulation. The solid red line represents the condition $\log_{10}\xi^{+}_\cPX=\log_{10}\xi_\cPX$.}
\label{fig:5a}
\end{figure}

Since part of the hadronic system X escapes the detector through the forward beam hole, and since low-energy particles remain undetected because of the PF object thresholds, the reconstructed $\xi^{+}_\cPX$ values are underestimated. This can be seen in Fig.~\ref{fig:5a}, which shows a scatter plot of reconstructed vs. generated values of $\xi_\cPX$ for \PYTHIAMBR events in the FG2 sample. As $\xi_\cPX$ decreases, its unmeasured fraction increases (the beam hole size is fixed), resulting in a larger deviation from the $\log_{10}\xi^{+}_\cPX=\log_{10}\xi_\cPX$ line. The calibration factor $C(\xi^{+}_\cPX)$, which brings the reconstructed values of $\xi^{+}_\cPX$ (Eq.~(\ref{eq:xirecodef})) to their true values (Eq.~(\ref{eq:xidef})) according to the formula $\log_{10}\xi_\cPX^\text{cal}=\log_{10}\xi^{+}_\cPX+C(\xi^{+}_\cPX)$, is evaluated from the \PYTHIAMBR simulation, by studying the $\log_{10}\xi_\cPX - \log_{10}\xi^{+}_\cPX$ difference in bins of $\log_{10}\xi^{+}_\cPX$. The factor $C(\xi^{+}_\cPX)$ decreases from the value of 1.1 at $\log_{10}\xi^{+}_\cPX \approx -6.5$ to 0.2 at $\log_{10}\xi^{+}_\cPX \approx -2.5$, with an uncertainty of 9\%, estimated by comparing the \PYTHIAMBR and \PYTHIA8~4C simulations.

\begin{figure*}[thb]
\centering
\includegraphics[width=\textwidth]{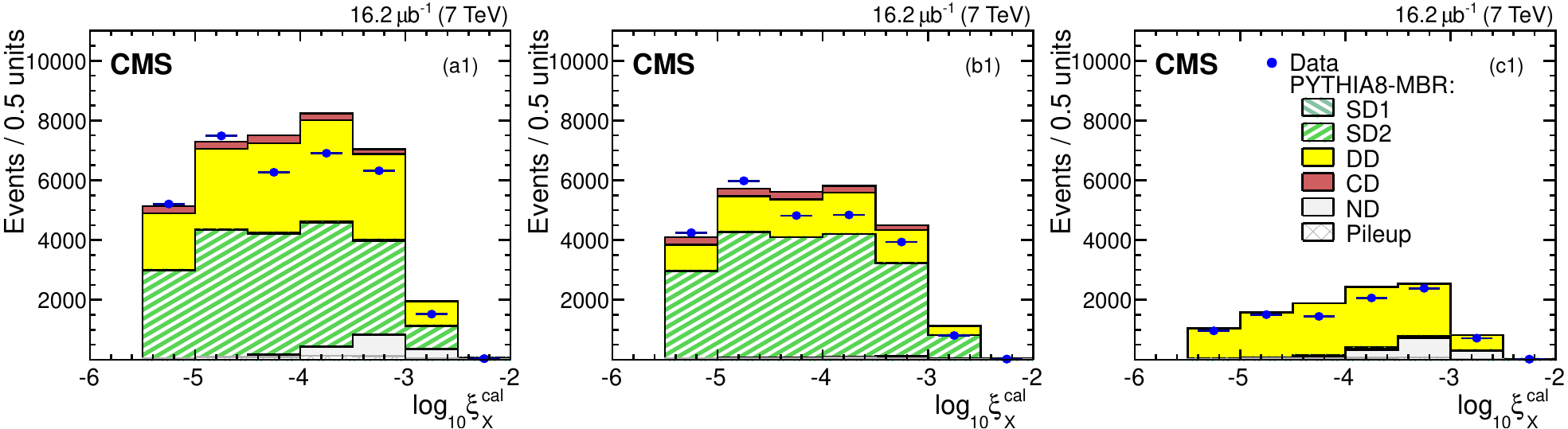}
\includegraphics[width=\textwidth]{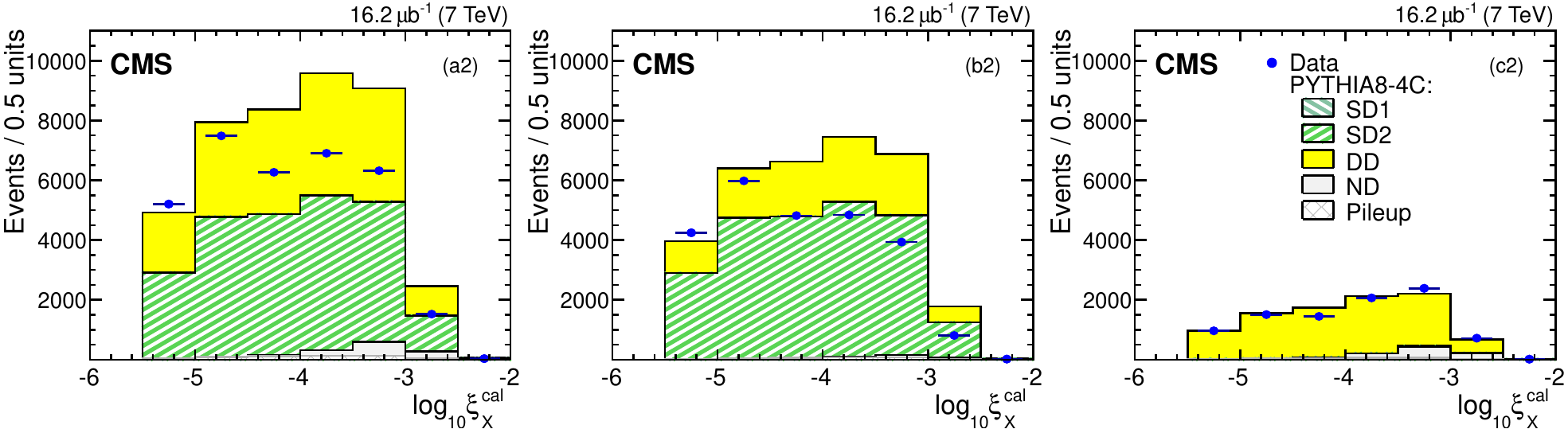}
\caption{Detector-level distributions of the reconstructed and calibrated $\xi_\cPX$ for (a) the entire FG2 sample, and the FG2 subsamples with (b) no CASTOR tag, and (c) a CASTOR tag (statistical errors only). The data are compared to the predictions of the \PYTHIAMBR (top three plots) and \PYTHIA8~4C (bottom three plots) simulations, which are normalized to the integrated luminosity of the data sample. The contribution of each of the generated processes is shown separately.}
\label{fig:5}
\end{figure*}

Figure~\ref{fig:5} presents the distribution of the calibrated $\log_{10}\xi_\cPX^\text{cal}$ for the FG2 sample, compared to the predictions of the \PYTHIAMBR (top) and \PYTHIA8~4C (bottom) simulations. Figure~\ref{fig:5}a shows the comparison for the entire FG2 sample. The separation of the SD and DD processes (hatched green and solid yellow histograms, respectively) by means of the CASTOR tag is clearly seen in Figs.~\ref{fig:5}b,c. Overall, the \PYTHIAMBR MC describes the data better than \PYTHIA8~4C and is therefore used to extract the diffractive cross sections. Both MC predictions are presented for a Pomeron trajectory with $\varepsilon=0.08$ and describe the region of low $\xi_\cPX$ well. At higher $\xi_\cPX$ values, $\varepsilon=0.104$ would be more appropriate~\cite{Covolan:1996uy}, providing a better agreement with the data in that region. Since the $M_\cPX$ dependence of $\varepsilon$ is currently not available in the MC models used for this analysis, we extract cross sections using $\varepsilon=0.08$ and evaluate a systematic uncertainty related to the possible variation of $\varepsilon$, as explained in Section~\ref{sec:sdddsig}.

The differential cross sections measured in bins of $\xi_\cPX$, separately for $\log_{10}M_\cPY<0.5$ (no CASTOR tag) and $0.5<\log_{10}M_\cPY<1.1$ (CASTOR tag), are calculated with the formula
\begin{equation}
\frac{\rd\sigma}{\rd\log_{10}\xi_\cPX} = \frac{N^\text{evt}}{\lumi \, (\Delta \log_{10}\xi_\cPX)_\text{bin}},
\label{eq:xsecSD}
\end{equation}
where $N^\text{evt}$ is the number of events in the bin, corrected for acceptance and migration effects, $\lumi$ is the integrated luminosity, and $(\Delta \log_{10}\xi_\cPX)_\text{bin}$ is the bin width. The acceptance and migration corrections are evaluated with the iterative Bayesian unfolding technique~\cite{D'Agostini:1994zf}, as implemented in the \textsc{RooUnfold} package~\cite{2011arXiv1105.1160A}, with four iterations. The number of iterations is optimized following the procedure suggested in Ref.~\cite{D'Agostini:1994zf}, by studying the difference in $\chi^2$ (goodness-of-fit) values after consecutive iterations; the final unfolded distribution, folded back to the detector level, is consistent with the observed data. The response matrix is obtained with \PYTHIAMBR ($\varepsilon = 0.08$). The results are presented in Table~\ref{tab:1}; they include a small correction for overlapping $\Pp\Pp$ collisions ($\sim$7\%), evaluated by comparing MC simulations with and without pileup.

\begin{table*}[t!h]
\centering
\topcaption{The differential forward pseudorapidity gap cross sections $\rd\sigma/\rd\log_{10}\xi_\cPX$ for $\log_{10}M_\cPY<0.5$ (SD dominated, without CASTOR tag) and $0.5<\log_{10}M_\cPY<1.1$ (DD dominated, with CASTOR tag). The first and the second errors correspond to statistical and systematic uncertainties, respectively.}
\label{tab:1}
\renewcommand{\arraystretch}{1.5}
\begin{scotch}{ccc}
bin & $\rd\sigma_\text{no-CASTOR}/\rd\log\xi_\cPX$ (mb) & $\rd\sigma_\text{CASTOR}/\rd\log\xi_\cPX$  (mb)\\
\hline
 $ -5.5 <\log_{10}\xi_\cPX < -5.0$ & $1.17 \pm 0.02_{-0.11}^{+0.08}$ & $0.30 \pm 0.01_{-0.04}^{+0.03}$\\
 $ -5.0 <\log_{10}\xi_\cPX < -4.5$ & $1.16 \pm 0.02_{-0.17}^{+0.18}$ & $0.30 \pm 0.01 \pm 0.04$\\
 $ -4.5 <\log_{10}\xi_\cPX < -4.0$ & $0.91 \pm 0.02_{-0.12}^{+0.15}$& $0.26 \pm 0.01 \pm 0.03$\\
 $ -4.0 <\log_{10}\xi_\cPX < -3.5$ & $0.88 \pm 0.02_{-0.09}^{+0.10}$& $0.32 \pm 0.01_{-0.05}^{+0.03}$\\
 $ -3.5 <\log_{10}\xi_\cPX < -3.0$ & $0.98 \pm 0.02_{-0.13}^{+0.14}$& $0.51 \pm 0.01_{-0.05}^{+0.06}$\\
 $ -3.0 <\log_{10}\xi_\cPX < -2.5$ & $0.78 \pm 0.03_{-0.09}^{+0.11}$& $0.67 \pm 0.03_{-0.10}^{+0.12}$\\
\end{scotch}

\end{table*}

\begin{figure*}[tbh]
\centering
\includegraphics[width=\textwidth]{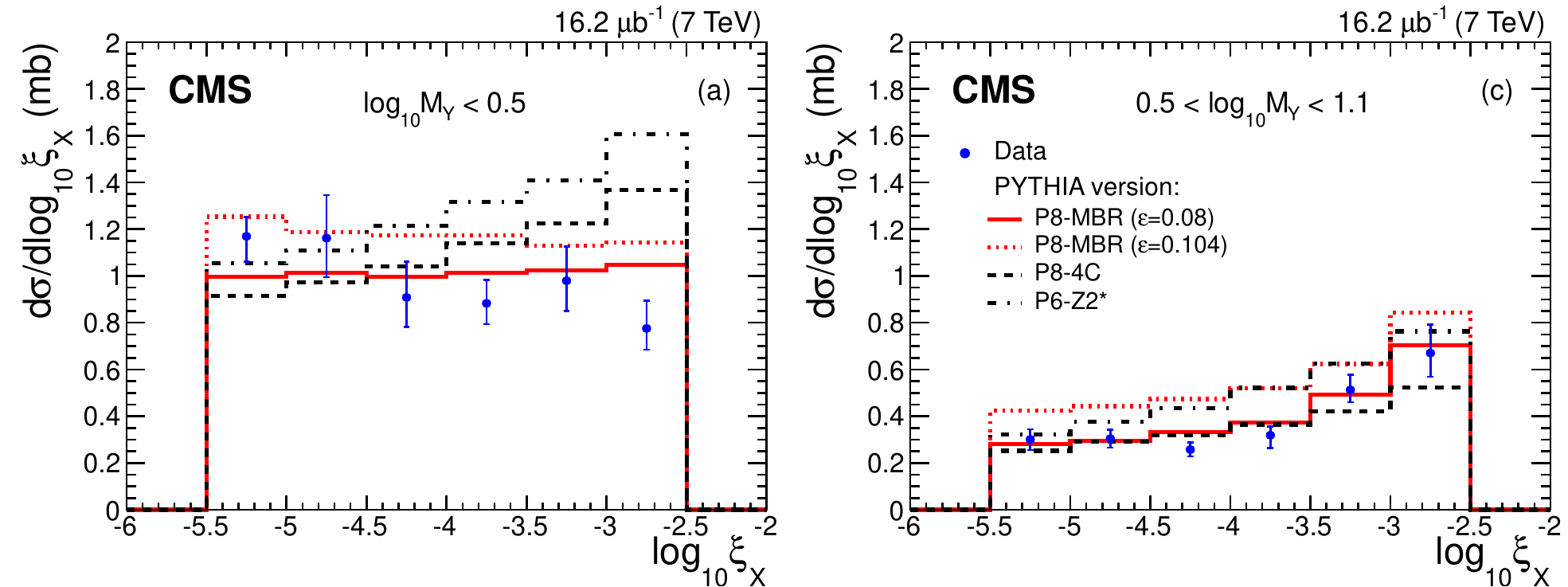}
\includegraphics[width=\textwidth]{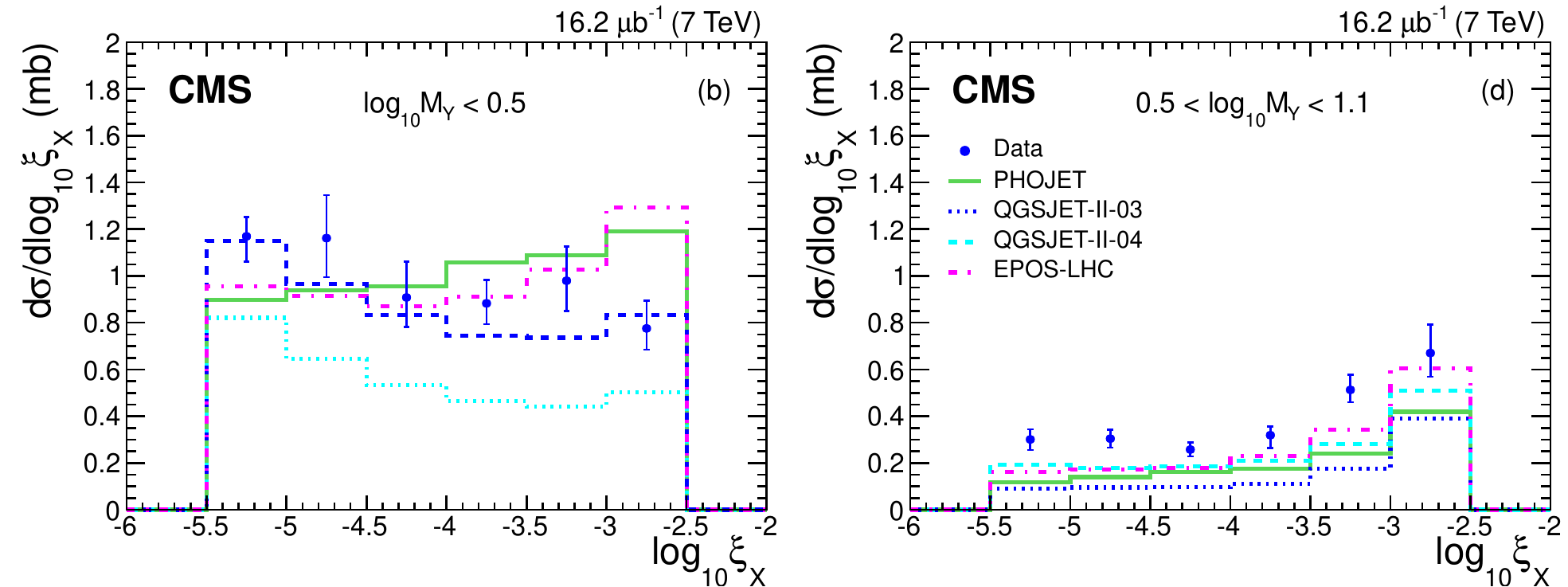}
\caption{Cross sections $\rd\sigma/\rd\log_{10}\xi_\cPX$ for (a,b) $\log_{10}M_\cPY<0.5$ (SD dominated) and (c,d) $0.5<\log_{10}M_\cPY<1.1$ (DD dominated) compared to MC predictions: (a,c) \PYTHIAMBR, \PYTHIA8~4C, \PYTHIA6~Z2*, and (b,d) \PHOJET, \QGSJETII~03, \QGSJETII~04, \EPOS. Error bars are dominated by systematic uncertainties (discussed in Section~\ref{sec:syst}).}
\label{fig:6}
\end{figure*}

Figures~\ref{fig:6}a,c present the measured cross sections compared to the \PYTHIAMBR, \PYTHIA8~4C, and \PYTHIA6~Z2* predictions. The error bars of the data points are dominated by systematic uncertainties, which are discussed in Section~\ref{sec:syst}. The predictions of \PYTHIAMBR are shown for two values of the $\varepsilon$ parameter of the Pomeron trajectory. Both values, $\varepsilon = 0.08$ and $\varepsilon = 0.104$, describe the measured cross section for $\log_{10}M_\cPY<0.5$. The data for $0.5<\log_{10}M_\cPY<1.1$ favor the smaller value of $\varepsilon$, specifically in the region of lower $\xi_\cPX$, corresponding to the topology in which both dissociation masses are low. The prediction of the Schuler-Sj\"ostrand model, used in the \PYTHIA8~4C simulation, describes well the measured cross section for $0.5<\log_{10}M_\cPY<1.1$, while the \PYTHIA6~Z2* simulation overestimates it. Both predictions are higher than the data for $\log_{10}M_\cPY<0.5$ at high $\log_{10}\xi_\cPX$, and the predicted rising behavior of the cross section is not confirmed by the data in the region of the measurement, $-5.5<\log_{10}\xi_\cPX<-2.5$.

Figures~\ref{fig:6}b,d present a comparison of the measured cross sections with the \PHOJET, \QGSJETII~03, \QGSJETII~04, and \EPOS predictions. None of the models is able to describe the magnitude of the cross section in the region $0.5<\log_{10}M_\cPY<1.1$. For $\log_{10}M_\cPY<0.5$, the \PHOJET and \EPOS generators fail to describe the falling behavior of the data, \QGSJETII~03 describes the measured cross section reasonably well, while \QGSJETII~04 underestimates the magnitude of the cross section.

\section{Central pseudorapidity gap cross section from the CG event sample}
\label{sec:dd}

The cross section for events with a central pseudorapidity gap is measured as a function of the variable $\Delta\eta$, defined as $\Delta\eta = - \log \xi$, where $\xi = M^2_\cPX \, M^2_\cPY/(s \, m^2_\Pp)$, with $m_\Pp$ the proton mass. For purely DD events, the position of the gap center is related to the dissociation masses by the expression $\eta_\mathrm{c} = \log(M_\cPY/M_\cPX)$.

\begin{figure}[tbh]
\centering
\includegraphics[width=0.48\textwidth]{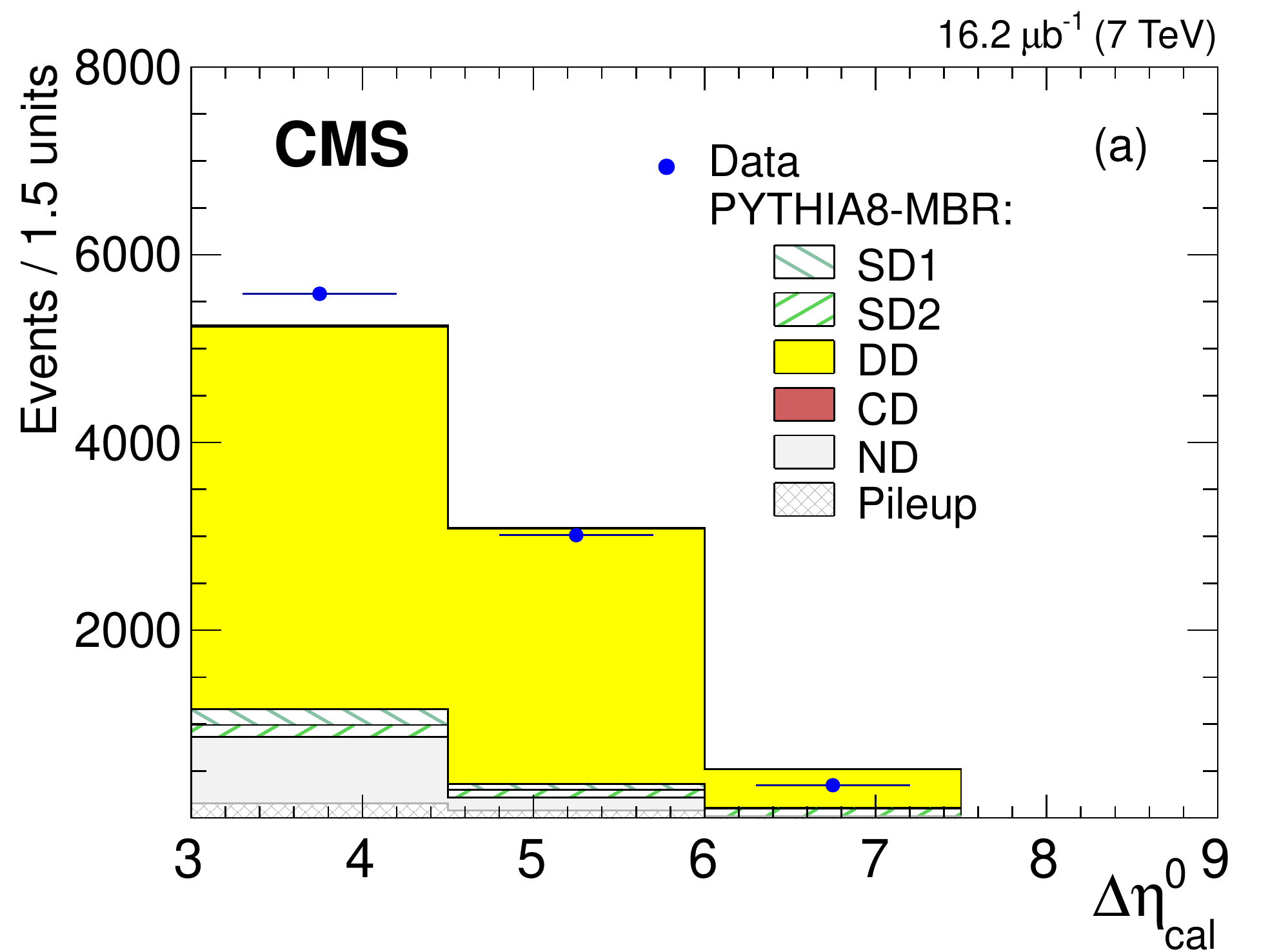}
\includegraphics[width=0.48\textwidth]{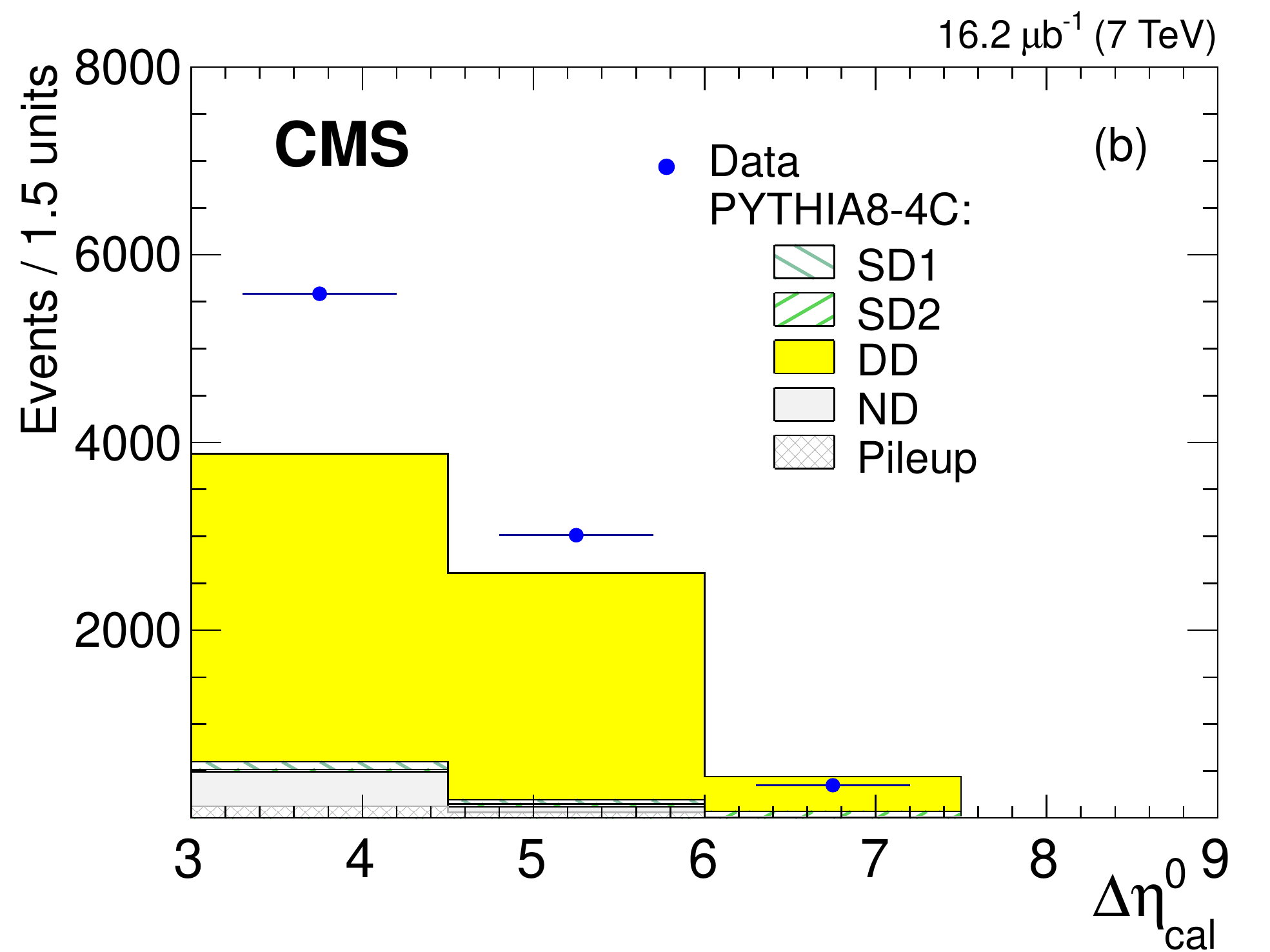}
\caption{Detector-level distributions of reconstructed and calibrated $\Delta\eta^{0}$ values for the measured CG sample with a central LRG. The data are compared to predictions of (a) \PYTHIAMBR, and (b) \PYTHIA8~4C simulations normalized to the integrated luminosity of the data sample. Contributions for each of the generated processes are shown separately.}
\label{fig:70}
\end{figure}

As discussed in Section~\ref{sec:topo}, the central-gap width (Fig.~\ref{fig:3}c) is reconstructed as $\Delta\eta^{0} = \eta^{0}_\text{max}-\eta^{0}_\text{min}$.  The calibration factor $C$, which corrects $\Delta\eta^{0}$ for detector effects according to the formula $\Delta\eta^{0}_\mathrm{cal} = \Delta\eta^{0}_\mathrm{rec}-C$, is extracted from the \PYTHIAMBR MC as the difference $C = \Delta\eta^{0}_\mathrm{rec}-\Delta\eta^{0}_\mathrm{gen}$. It amounts to $C=2.42 \pm 0.12$, with the uncertainty estimated from a comparison with \PYTHIA8~4C. Figure~\ref{fig:70} presents the distribution of the calibrated $\Delta\eta^{0}$ for the CG sample along with simulated distributions from \PYTHIAMBR and \PYTHIA8~4C.

The differential cross section, measured in bins of $\Delta\eta$ for $\Delta\eta>3$, $\log_{10}M_\cPX>1.1$, and $\log_{10}M_\cPY>1.1$, is calculated according to the formula
\begin{equation}
\frac{\rd\sigma}{\rd\Delta\eta} = \frac{N^\text{evt}}{\lumi \, (\Delta \eta)_\text{bin}},
\label{eq:xsecDD2}
\end{equation}
where $N^\text{evt}$ is the number of events in a given bin, corrected for acceptance and migration effects, and also for the extrapolation from $\Delta\eta^{0}>3$ (for gaps overlapping $\eta=0$) to $\Delta\eta>3$ for all gaps, $\lumi$ is the integrated luminosity, and $(\Delta \eta)_\text{bin}$ is the bin width. The acceptance and migration corrections are evaluated with the iterative Bayesian unfolding technique~\cite{D'Agostini:1994zf} with two iterations, optimized as described in Section~\ref{sec:sd}. The response matrix is obtained using \PYTHIAMBR with $\varepsilon = 0.08$.

\begin{table}[bht]
\centering
\topcaption{The differential central pseudorapidity gap (DD dominated) cross section $\rd\sigma/\rd\Delta\eta$ for $\Delta \eta >3$, $\log_{10}M_\cPX>1.1$, and $\log_{10}M_\cPY>1.1$. The first and second errors correspond to statistical and systematic uncertainties, respectively.}
\label{tab:2}
\renewcommand{\arraystretch}{1.5}
\begin{scotch}{cc}
 $\Delta\eta$ bin & $\rd\sigma_\mathrm{CG}/\rd\Delta\eta$ (mb)\\
\hline
 $ 3.0 <\Delta \eta < 4.5$ & $0.25 \pm 0.003_{-0.04}^{+0.05}$ \\
 $ 4.5 <\Delta \eta < 6.0$ & $0.11 \pm 0.002_{-0.02}^{+0.03}$\\
 $ 6.0 <\Delta \eta < 7.5$ & $0.032 \pm 0.001 \pm 0.009$\\
\end{scotch}
\end{table}

\begin{figure*}[tbh]
\centering
\includegraphics[width=0.49\textwidth]{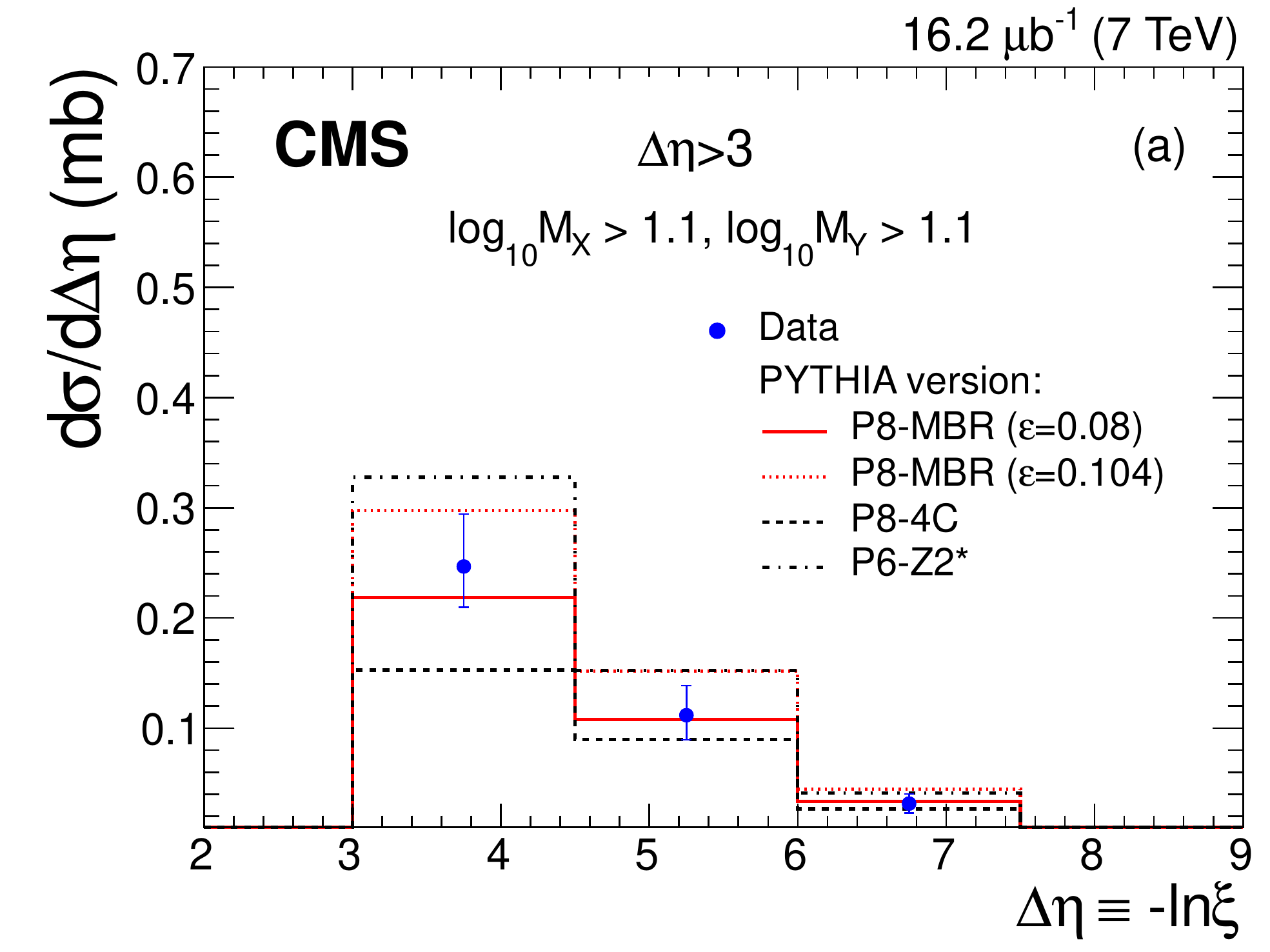}
\includegraphics[width=0.49\textwidth]{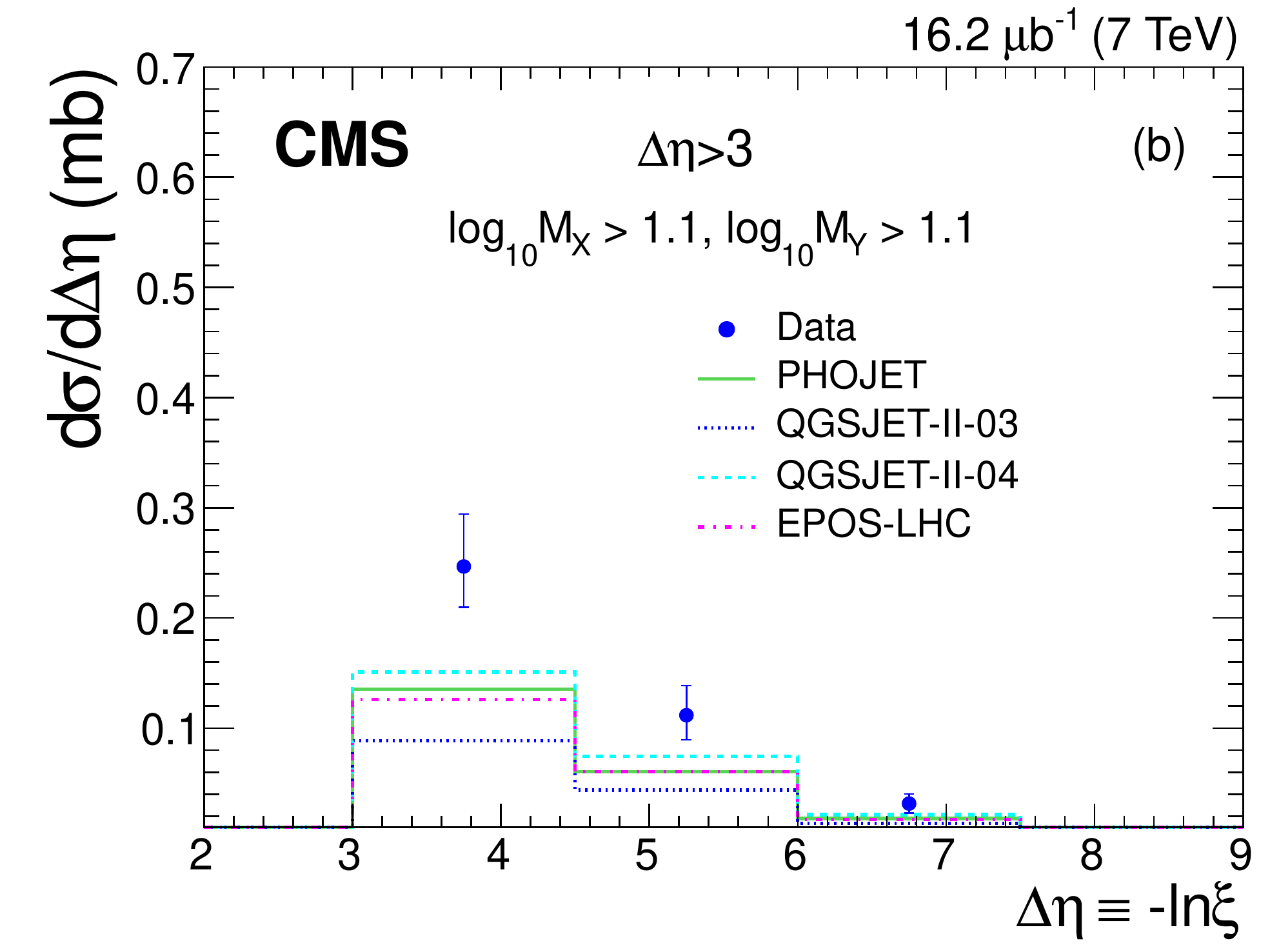}
\caption{The central pseudorapidity-gap cross section $\rd\sigma/\rd\Delta\eta$ (DD dominated) compared to MC predictions: (a) \PYTHIAMBR, \PYTHIA8~4C, and \PYTHIA6~Z2*, and (b) \PHOJET, \QGSJETII~03, \QGSJETII~04, and \EPOS. Error bars are dominated by systematic uncertainties, which are discussed in Section~\ref{sec:syst}.}
\label{fig:7}

\end{figure*}

The measured differential cross section is presented in Table~\ref{tab:2}, and compared to predictions of theoretical models in Fig.~\ref{fig:7}. The results take into account the pileup correction, and the uncertainties are dominantly systematic (see Section~\ref{sec:syst}). The prediction of the \PYTHIAMBR MC simulation with both $\varepsilon= 0.08$ and 0.104 describes well the central-gap data. \PYTHIA8~4C underestimates the data in all bins, while \PYTHIA6~Z2* overestimates the data in the lowest $\Delta \eta$ bin. The \PHOJET, \QGSJETII~03, \QGSJETII~04, and \EPOS generators underestimate the magnitude of the measured cross section.

\section{Integrated cross sections}
\label{sec:integsig}

The forward and central pseudorapidity gap samples are also used to measure the integrated cross sections in the kinematic regions given in Sections~\ref{sec:sd} and \ref{sec:dd}. The forward pseudorapidity-gap cross sections, $\sigma_\text{no-CASTOR}$ and $\sigma_\text{CASTOR}$, are measured in the region $-5.5<\log_{10}\xi_\cPX<-2.5$, for events without and with a CASTOR tag, corresponding to $\log_{10}M_\cPY<0.5$, and $0.5<\log_{10}M_\cPY<1.1$, respectively, while the central pseudorapidity-gap cross section, $\sigma_\mathrm{CG}$, is measured for $\Delta\eta>3$, $\log_{10}M_\cPX>1.1$, and $\log_{10}M_\cPY>1.1$.

Each cross section is evaluated by means of the formula
\begin{equation}
\sigma = \frac{N^\text{evt}}{A \,\lumi},
\label{eq:xsec1bin}
\end{equation}
where $N^\text{evt}$ is the number of events in the kinematic regions given above, $A$ is the acceptance, defined as the ratio of the number of events reconstructed to the number of events generated in that bin, taking into account the pileup correction, and $\lumi$ is the integrated luminosity. The acceptance is evaluated with the \PYTHIAMBR MC generator.

Values of $\sigma_\text{no-CASTOR} = 2.99 \pm 0.02\stat{}_{-0.29}^{+0.32}\syst\unit{mb}$, $\sigma_\text{CASTOR} = 1.18 \pm 0.02\stat \pm 0.13\syst$\unit{mb}, and $\sigma_\mathrm{CG}=0.58 \pm 0.01\stat{}_{-0.11}^{+0.13}\syst$\unit{mb} are obtained. Systematic uncertainties are evaluated as discussed in Section~\ref{sec:syst}. As a consistency check of the analysis procedure, we measure the part of the total inelastic cross sections that is visible in the central CMS detector, corresponding to the region $\log_{10}M_\cPX>1.1$ or  $\log_{10}M_\cPY>1.1$. A value of $\sigma_\mathrm{vis}^\mathrm{check}= 61.29 \pm 0.07\stat$\unit{mb} is found, in good agreement with the published CMS result $\sigma_\mathrm{inel}(\xi>5 \times 10^{-6}) = 60.2 \pm 0.2\stat \pm 1.1\syst \pm 2.4\lum$\unit{mb}, measured in a slightly different kinematic region ($M_\cPX$ or $M_\cPY\gtrsim 16.7$\GeV)~\cite{Chatrchyan:2012nj}. According to \PYTHIAMBR, the phase space difference between the two measurements corresponds to $\sigma_\mathrm{vis}^\mathrm{check}-\sigma_\mathrm{inel}(\xi>5 \times 10^{-6})=0.5$\unit{mb}.

\begin{table*}[tbh]
\centering
\topcaption{Measured $\sigma_\text{no-CASTOR}$, $\sigma_\text{CASTOR}$, and $\sigma_\mathrm{CG}$ cross sections, compared to predictions of MC models. The first and the second errors in the data correspond to statistical and systematic uncertainties, respectively.}
\label{tab:3}
\begin{scotch}{cccc}
Cross section & $\sigma_\text{no-CASTOR}$(mb) & $\sigma_\text{CASTOR}$ (mb)& $\sigma_\mathrm{CG}$ (mb)\\
& SD dominated & DD dominated & DD dominated\\
\hline
 Data & $2.99 \pm 0.02_{-0.29}^{+0.32}$& $1.18 \pm 0.02 \pm 0.13$ & $0.58 \pm 0.01_{-0.11}^{+0.13}$ \rule[-2mm]{0pt}{7mm}\\
\hline
 \PYTHIAMBR & 3.05 & 1.24 & 0.54\\
 \PYTHIA8~4C  & 3.31 & 1.10 & 0.40\\
 \PYTHIA6~Z2* & 3.86 & 1.52 & 0.78\\
\hline
 \PHOJET      & 3.06 & 0.63 & 0.32\\
 \QGSJETII~03 & 2.63 & 0.48 & 0.22\\
 \QGSJETII~04 & 1.70 & 0.78 & 0.37\\
 \EPOS        & 2.99 & 0.85 & 0.31\\
\end{scotch}

\end{table*}

Table~\ref{tab:3} lists the measured cross sections together with the absolute predictions of the MC simulations. Based on the results presented thus far, the following conclusions can be drawn about the models: \PHOJET, \QGSJETII~03, \QGSJETII~04, and \EPOS predict too few DD events, which dominate the measured $\sigma_\text{CASTOR}$ and $\sigma_\mathrm{CG}$ cross sections (Figs. \ref{fig:6}d and \ref{fig:7}b); among these four models only \QGSJETII~03 satisfactorily predicts the $\sigma_\text{no-CASTOR}$ cross section (Fig.~\ref{fig:6}b); \PYTHIA8~4C, \PYTHIA6~Z2*, \PHOJET, and \EPOS do not predict correctly the $\xi_\cPX$ dependence for the SD process, which dominates the measured forward pseudorapidity gap cross section for $\log_{10}M_\cPY<0.5$ (Fig.~\ref{fig:6} a,b); and \PYTHIAMBR describes the data within uncertainties in all the measured regions.

\section{The SD and DD cross sections}
\label{sec:sdddsig}

The $\sigma_\text{no-CASTOR}$ cross section discussed above is dominated by SD events (Fig.~\ref{fig:5}b), whereas the $\sigma_\text{CASTOR}$ and $\sigma_\mathrm{CG}$ cross sections are mainly due to DD events (Figs.~\ref{fig:5}c and \ref{fig:70}). As the contribution from ND and other diffractive processes to these cross sections is small, we use the event decomposition as defined in the \PYTHIAMBR simulation with $\varepsilon = 0.08$ to correct for them and extract the SD and DD cross sections.

The dominant background in the $\sigma_\text{no-CASTOR}$ cross section originates from DD events; the CD contribution is minimal, while the ND contribution is negligible (Fig.~\ref{fig:5}b). The DD contribution is well understood via the CASTOR-tag events (Fig.~\ref{fig:5}c), and has an uncertainty of $\sim$10--20\% due to the ND contamination. Since the DD events contribute about 20\% to the no-CASTOR-tag sample, the uncertainty in the SD cross section due to the subtraction of the DD component amounts to only a few percent. The visible part of the total SD cross section, corresponding to $-5.5<\log_{10}\xi_\cPX<-2.5$, is found to be $\sigma^\mathrm{SDvis} = 4.06 \pm 0.04\stat _{-0.63}^{+0.69}\syst$\unit{mb}. The result accounts for both $\Pp\Pp\to \cPX\Pp$ and $\Pp\Pp\to \Pp \cPY$.

The dominant background to the $\sigma_\text{CASTOR}$ and $\sigma_\mathrm{CG}$ cross sections originates from ND events. The CD and SD contributions are negligible in the CASTOR-tag sample, while SD events contribute minimally to the central-gap sample (Figs.~\ref{fig:5}c and \ref{fig:70}a). The total DD cross sections integrated over the regions
\begin{itemize}
\item $-5.5<\log_{10}\xi_\cPX<-2.5$ and $0.5<\log_{10}M_\cPY<1.1$, and
\item $\Delta\eta>3$, $\log_{10}M_\cPX>1.1$ and $\log_{10}M_\cPY>1.1$,
\end{itemize}
are $\sigma^\mathrm{DDvis}_\text{CASTOR} = 1.06 \pm 0.02\stat \pm 0.12\syst$\unit{mb}, and $\sigma^\mathrm{DDvis}_\mathrm{CG} = 0.56 \pm 0.01 \stat {}_{-0.13}^{+0.15} \syst$\unit{mb}, respectively.

To provide the DD cross section in the widest kinematic region spanned by the data, we also evaluate the visible DD cross section, $\sigma^\mathrm{DDvis}$, defined as $\sigma^\mathrm{DDvis} = 2\,\sigma^\mathrm{DDvis}_\text{CASTOR}+\sigma^\mathrm{DDvis}_\mathrm{CG}$, where the factor of 2 assumes the same dependence of the DD cross section on $M_\cPX$ and $M_\cPY$ (boxed regions in Fig.~\ref{fig:4}b). This leads to $\sigma^\mathrm{DDvis} = 2.69 \pm 0.04 \stat {}_{-0.30}^{+0.29} \syst$,  in the kinematic region delimited by the solid and dashed (red) lines in Figs.~\ref{fig:4}b,c. This result is used below to extrapolate the DD cross section to the region $\Delta\eta>3$ (to the left of the dashed (blue) line in Fig.~\ref{fig:4}c).

\subsection{Extrapolation of the visible SD and DD cross sections}
\label{sec:sdddextrap}

The measurements based on the central CMS detector are insensitive to the low-mass part of diffractive dissociation. Therefore, in order to compare the measured $\sigma^\mathrm{SDvis}$ cross section with results of other experiments and theoretical models that present integrated cross sections for $\xi<0.05$, an extrapolation from $-5.5<\log_{10}\xi_\cPX<-2.5$ ($\xi_\cPX=M^2_\cPX/s$) to $\xi<0.05$ is required. Similarly, the $\sigma^\mathrm{DDvis}$ cross section must be extrapolated to $\Delta\eta>3$ (Fig.~\ref{fig:4}). The extrapolation factors, calculated by using each of the MC simulations introduced in Section~\ref{sec:mc}, are presented in Appendix A. Not all of the simulations are able to describe the measured cross sections (see Section~\ref{sec:integsig}), nor do they include realistic hadronization models (see  Appendix B). Following the discussion in Section~\ref{sec:integsig} and Appendix B, the extrapolation factors are determined with \PYTHIAMBR (with $\varepsilon = 0.08$), which describes well all aspects of our data. The multiplicative factor needed to extrapolate the measured SD cross section to $\xi<0.05$ is $f^\mathrm{SD}_\mathrm{MBR}=2.18^{+13\%}_{-4\%}$, and that for the extrapolation of the DD cross section to $\Delta\eta>3$ is $f^\mathrm{DD}_\mathrm{MBR}=1.92^{+31\%}_{-10\%}$ (Tables~\ref{tab:fsd} and \ref{tab:fdd} in Appendix A). The extrapolation uncertainties are estimated by changing the parameters $\alpha'$ and $\varepsilon$ of the Pomeron trajectory from their nominal values ($\alpha'= 0.25\GeV^{-2},~\varepsilon = 0.08$) to those presented in Tables~\ref{tab:fsd} and \ref{tab:fdd} (one parameter changed at a time), and adding in quadrature the corresponding deviations with respect to the central result, separately for the positive and negative deviations. The extrapolated SD and DD cross sections thus obtained are $\sigma^\mathrm{SD}=8.84 \pm 0.08 \stat {}^{+1.49}_{-1.38} \syst ^{+1.17}_{-0.37}\,\text{(extrap)}$\unit{mb} and
$\sigma^\mathrm{DD} = 5.17 \pm 0.08 \stat {}^{+0.55}_{-0.57} \syst ^{+1.62}_{-0.51}\,\text{(extrap)}$\unit{mb}, respectively.

\begin{figure}[tbh]
\centering
\includegraphics[width=\cmsFigWidth]{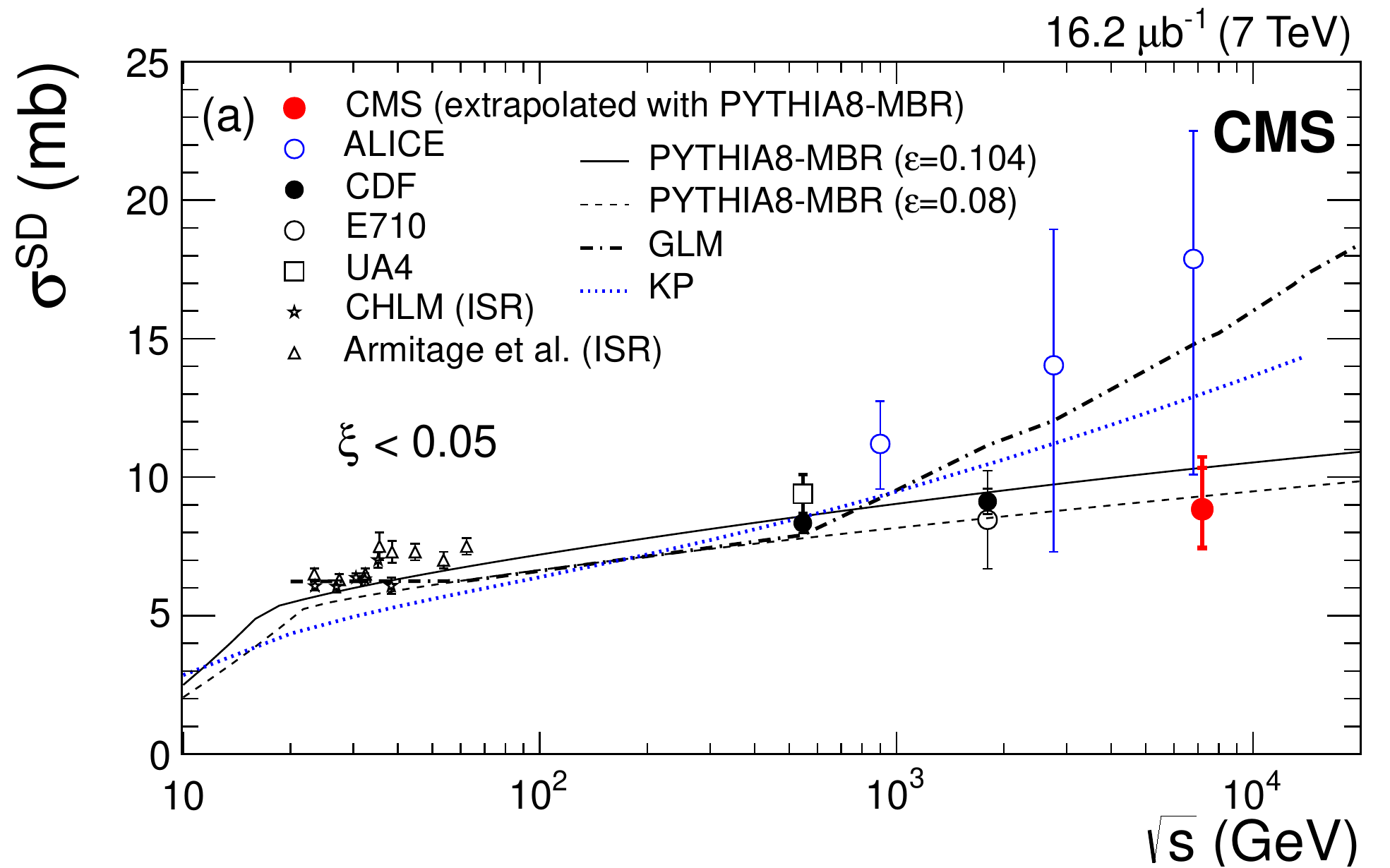}
\includegraphics[width=\cmsFigWidth]{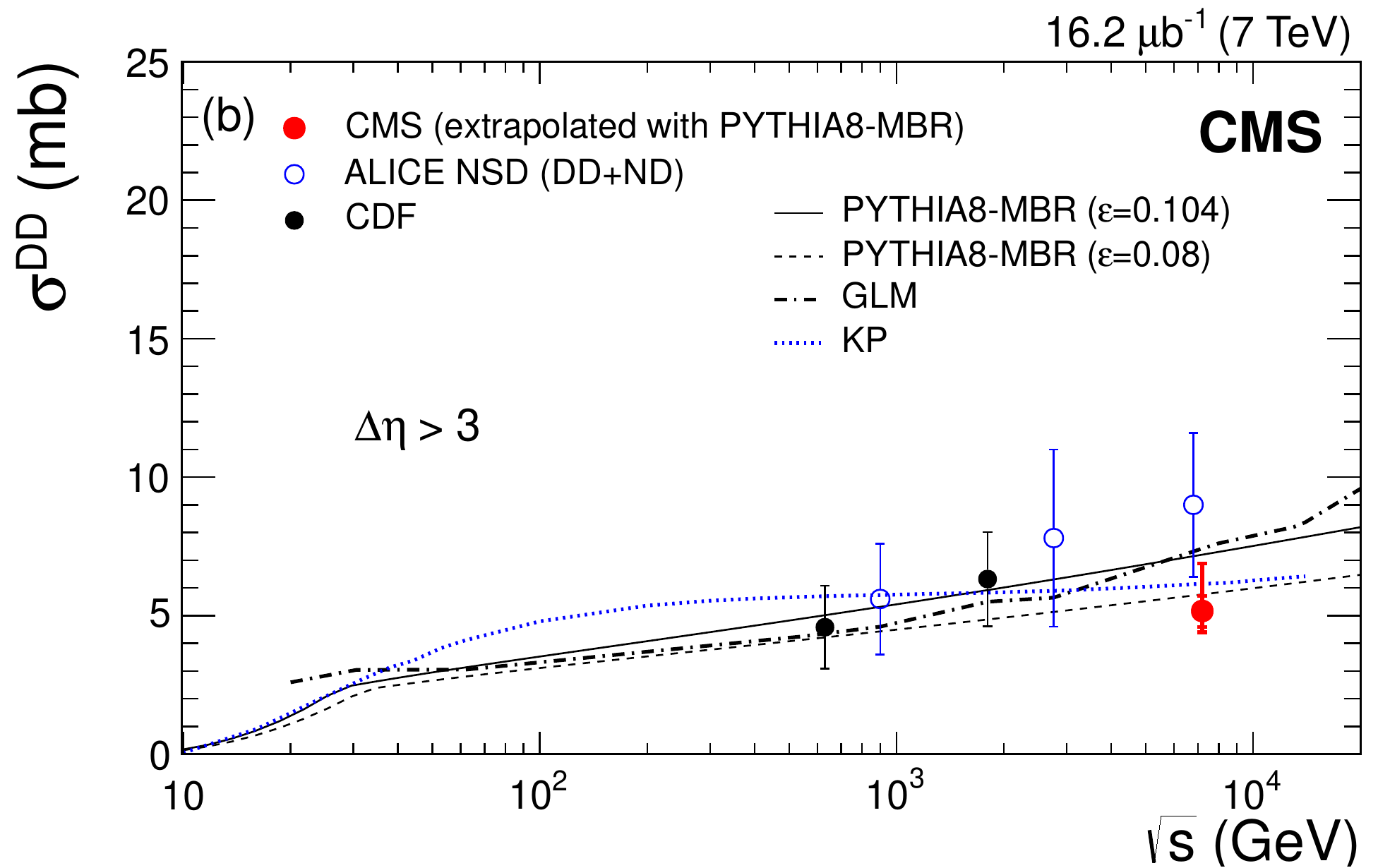}
\caption{Diffractive cross sections as a function of collision energy measured in $\Pp\Pp$ and $\Pp\Pap$ collisions~\cite{Abelev:2012sea,PhysRevD.50.5535,Amos:1992jw,Bernard:1986yh,Albrow:1976sv,1982NuPhB.194..365A,2001PhRvL..87n1802A} compared to \PYTHIAMBR ($\varepsilon = 0.08, 0.104$) and other model predictions~\cite{2012PhLB..716..425G,2010EPJC...67..397K,Kaidalov:2011bu}: (a) total SD cross section for $\xi<0.05$, and (b) total DD cross section for $\Delta\eta>3$. The inner (outer) error bars of the CMS data points correspond to the statistical and systematic (and the additional extrapolation) uncertainties added in quadrature.}
\label{fig:61}
\end{figure}

Figure~\ref{fig:61}a presents the extrapolated SD cross section compared to the ALICE result~\cite{Abelev:2012sea} and a compilation of lower center-of-mass energy measurements~\cite{PhysRevD.50.5535,Amos:1992jw,Bernard:1986yh,Albrow:1976sv,1982NuPhB.194..365A}. The data are also compared to the \PYTHIAMBR simulation, as well as to the GLM~\cite{2012PhLB..716..425G} and KP~\cite{2010EPJC...67..397K,Kaidalov:2011bu} models. The CMS result is consistent with a SD cross section weakly rising with energy.

Figure~\ref{fig:61}b shows the extrapolated DD cross section compared to the ALICE results~\cite{Abelev:2012sea}, those by CDF at $\sqrt{s}=630$\GeV and $1.8$\TeV~\cite{2001PhRvL..87n1802A}, as well as the \PYTHIAMBR,  GLM~\cite{2012PhLB..716..425G}, and KP \cite{2010EPJC...67..397K,Kaidalov:2011bu} models. The CMS measurement at $\sqrt{s}=7$\TeV is in agreement with the ALICE measurement at the same energy. Note, however, that the ALICE result is obtained from the NSD (non-single-diffractive = DD+ND) data, while for the CMS measurement the ND background has been subtracted. Here as well, the data are consistent with a weakly rising cross section with energy, as predicted by the models.

\subsection{Summary of results}

\begin{table*}[t]
\centering
\topcaption{Measured diffractive cross sections in regions of $M_\cPX$ (or $\xi_\cPX=M^2_\cPX/s$), $M_\cPY$ (or $\xi_\cPY=M^2_\cPY/s$), and $\Delta\eta$ ($\Delta\eta=-\log\xi$, where $\xi = M^2_\cPX \, M^2_\cPY/(s \, m^2_\Pp)$ for DD). The method used for the cross section extraction is indicated as LRG for calculations involving all events selected in the LRG samples, and as MBR for calculations that involve background subtraction or extrapolation based on the prediction of the \PYTHIAMBR simulation. The first and the second errors in the data correspond to the statistical and systematic uncertainties, respectively. For $\sigma^\mathrm{SD}$ and $\sigma^\mathrm{DD}$, the third errors correspond to the extrapolation uncertainties.}
\label{tab:4}
\begin{scotch}{lcccr}
Cross section & $M_\cPX$ or $\xi_\cPX$ range & $M_\cPY$ or $\xi_\cPY$ range & $\Delta\eta$ range & Result (mb)\rule[-2mm]{0pt}{6mm}\\
\hline
\multicolumn{5}{l}{LRG}\rule[-2mm]{0pt}{6mm}\\
$\sigma_\text{no-CASTOR}$ & $-5.5 < \log{\xi}_\cPX < -2.5$ & $\log_{10} M_\cPY < 0.5$         &\NA& $2.99 \pm 0.02 ^{+0.32}_{-0.29}$\\
$\sigma_\text{CASTOR}$ & $-5.5 < \log{\xi}_\cPX < -2.5$ & $0.5 < \log_{10}M_\cPY < 1.1$ &\NA& $1.18 \pm 0.02 \pm 0.13$\\
$\sigma_\mathrm{CG}$ & $\log_{10} M_\cPX > 1.1$            & $\log_{10} M_\cPY > 1.1$ & $\Delta\eta > 3$ & $0.58 \pm 0.01 ^{+0.13}_{-0.11}$ \\
\hline
\multicolumn{5}{l}{MBR}\rule[-2mm]{0pt}{7mm}\\
\multirow{2}{*}{$\sigma^\mathrm{SDvis}$} & $-5.5 < \log{\xi}_\cPX < -2.5$ & $M_\cPY = m_\Pp$ &\NA& \multirow{2}{*}{$4.06 \pm 0.04 ^{+0.69}_{-0.63}$} \\
 & $M_\cPX = m_\Pp$ & $-5.5 < \log{\xi}_\cPY < -2.5$ &\NA& \\
\rule[-2mm]{0pt}{7mm} $\sigma^\mathrm{DDvis}_\text{CASTOR}$ & $-5.5 < \log{\xi}_\cPX < -2.5$ & $0.5 < \log_{10} M_\cPY < 1.1$ &\NA& $1.06 \pm 0.02 \pm 0.12$\\
\rule[-2mm]{0pt}{7mm} $\sigma^\mathrm{DDvis}_\mathrm{CG}$ & $\log_{10} M_\cPX > 1.1 $ & $\log_{10} M_\cPY > 1.1$ & $\Delta\eta > 3$ & $0.56 \pm 0.01 ^{+0.15}_{-0.13}$\\
\multirow{3}{*}{$\sigma^\mathrm{DDvis}$} & $-5.5 < \log{\xi}_\cPX < -2.5$ & $0.5 < \log_{10} M_\cPY < 1.1$ &\NA \rule[-2mm]{0pt}{7mm} & \multirow{3}{*}{$2.69 \pm 0.04^{+0.29}_{-0.30}$}\\
 & $0.5 < \log_{10} M_\cPX < 1.1$ & $-5.5 < \log{\xi}_\cPY < -2.5$ &\NA& \\
 & $\log_{10} M_\cPX > 1.1 $ & $\log_{10} M_\cPY > 1.1$ & $\Delta\eta > 3$ & \\
\hline
\multicolumn{5}{l}{MBR}\rule[-2mm]{0pt}{7mm}\\
\multirow{2}{*}{$\sigma^\mathrm{SD}$} & $\xi_\cPX < 0.05$ & $M_\cPY = m_\Pp$ &\NA\rule[-1mm]{0pt}{4mm} & \multirow{2}{*}{$8.84 \pm 0.08 ^{+1.49}_{-1.38} {\ }^{+1.17}_{-0.37}$}\\
 & $M_\cPX = m_\Pp$ & $\xi_\cPY < 0.05$ &\NA& \\
\rule[-2mm]{0pt}{7mm} $\sigma^\mathrm{DD}$ &\NA&\NA& $\Delta\eta > 3$ & $5.17 \pm 0.08^{+0.55}_{-0.57} {\ }^{+1.62}_{-0.51}$ \\
\end{scotch}

\end{table*}

Table~\ref{tab:4} presents the summary of the cross section measurements illustrated in the previous sections, together with the kinematic region covered by each measurement. The method used for the cross section extraction (LRG or MBR) is given as well. The $\sigma_\text{no-CASTOR}$, $\sigma_\text{CASTOR}$, and $\sigma_\mathrm{CG}$ cross sections are measured from all the events passing the LRG selection (Section ~\ref{sec:integsig}). The $\sigma^\mathrm{SDvis}$, $\sigma^\mathrm{DDvis}_\text{CASTOR}$, and $\sigma^\mathrm{DDvis}_\mathrm{CG}$ cross sections are extracted from the latter ones by subtracting the background contribution from other processes as predicted by the \PYTHIAMBR simulation (Section~\ref{sec:sdddsig}). In addition, $\sigma^\mathrm{DDvis}$ is calculated from the combination of $\sigma^\mathrm{DDvis}_\text{CASTOR}$ and $\sigma^\mathrm{DDvis}_\mathrm{CG}$. Finally, the $\sigma^\mathrm{SD}$ and $\sigma^\mathrm{DD}$ cross sections (Section~\ref{sec:sdddextrap}) are calculated by extrapolating $\sigma^\mathrm{SDvis}$ and $\sigma^\mathrm{DDvis}$ to the region of lower diffractive masses using the mass dependence of the cross section predicted by \PYTHIAMBR.

{\section{Pseudorapidity gap cross section}\label{sec:rapgapxsec}

This section presents the results of an alternative approach to the study of diffractive events, in which the data are analyzed in terms of the widest pseudorapidity gap adjacent to the edge of the detector~\cite{Nurse:2011vt}. In each event, particles are first ordered in $\eta$, and the largest pseudorapidity gap, $\Delta\eta^\mathrm{F}$, is determined as $\Delta\eta^\mathrm{F}=\max(\abs{\eta_\text{min}-\eta^-}, \abs{\eta_\text{max}-\eta^+})$, where $\eta^{\pm}= {\pm}4.7$ are the detector edges in $\eta$, and $\eta_\text{max}$ ($\eta_\text{min}$) is the highest (lowest) $\eta$ of the PF objects in the event (see Fig.~\ref{fig:2}).

\begin{figure}[tbh]
\centering
\includegraphics[width=0.49\textwidth]{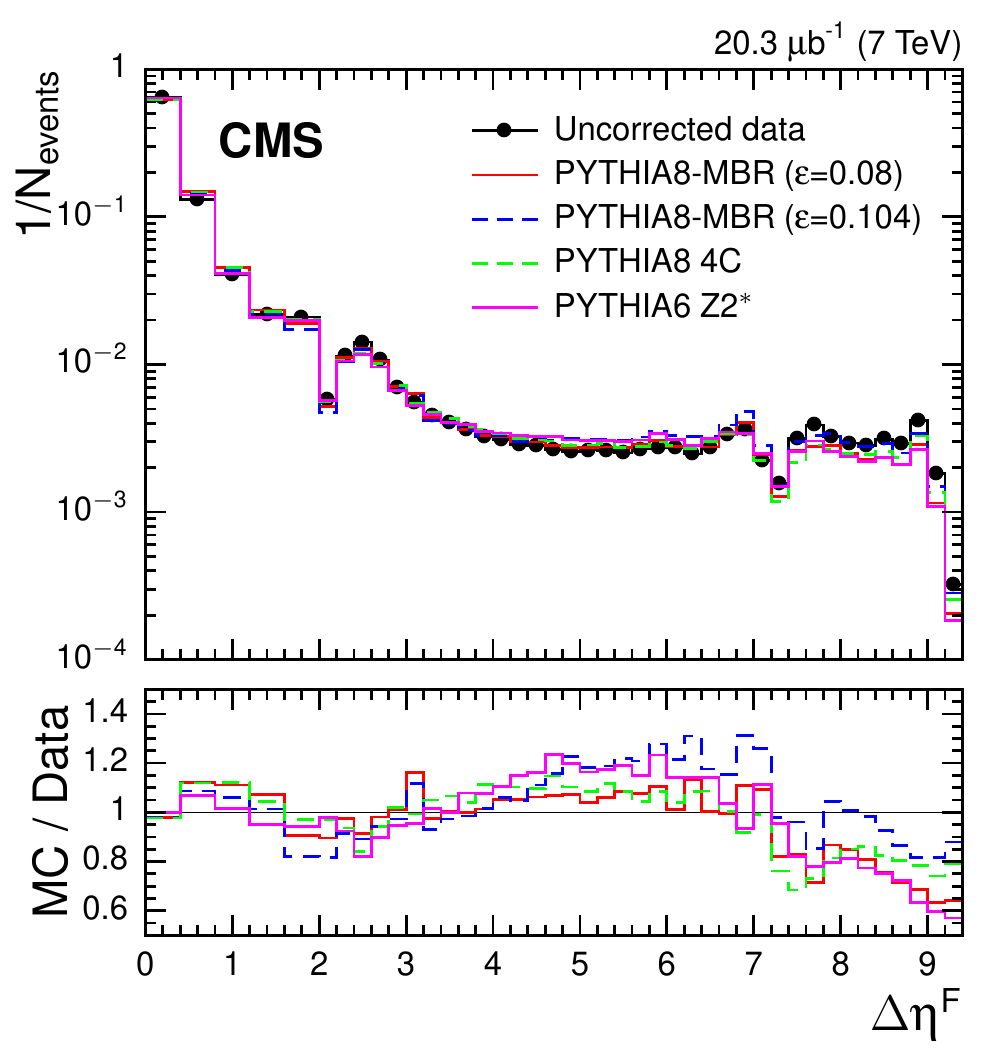}
\caption{Uncorrected $\Delta\eta^\mathrm{F}$ distribution compared to various detector-level MC predictions.}
\label{fig:raw}
\end{figure}

The analysis is based on a minimum bias data sample, selected as described in Section~\ref{sec:selection}, with negligible pileup (0.006), and corresponding to an integrated luminosity of 20.3\mubinv. The uncorrected distribution of the pseudorapidity gap size is shown in Fig.~\ref{fig:raw}, along with the predictions of various MC models. A wider bin width is used at low $\Delta\eta^\mathrm{F}$ to account for the lower spatial resolution in the forward region.

\subsection{Corrections for experimental effects}\label{sec:expcorrections}

Interactions of the beam protons with the residual gas particles in the beam pipe or inside the detector region affect the pseudorapidity gap distribution in data. The overall beam-induced background, integrated over the full measurement region, is about 0.7\%. After the subtraction of this background, the differential cross section $\rd\sigma / \rd\Delta\eta^\mathrm{F}$ is determined according to the formula

\begin{equation}
  \frac{\rd\sigma}{\rd\Delta\eta^\mathrm{F}} = \frac{N^\text{evt}}{T_{\varepsilon} \, \lumi\,  (\Delta\eta^\mathrm{F})_\text{bin}},
\end{equation}
where $N^\text{evt}$ is the number of events in the bin, corrected for migration effects, $T_{\varepsilon}$ the trigger efficiency, $\mbox{\lumi}$ the integrated luminosity, and $(\Delta\eta^\mathrm{F})_\text{bin}$ the bin width.

The trigger efficiency is obtained from a comparison with zero-bias data where no trigger requirements were applied. In order to have a satisfactory correlation between the generated and reconstructed values of $\Delta\eta^\mathrm{F}$, and hence a reliable correction for bin-migration effects, the cross section is evaluated for events with at least one stable final-state particle of transverse momentum $\pt > 200$\MeV in the region of $\abs{\eta}<4.7$. The migration corrections are evaluated with the iterative Bayesian unfolding technique~\cite{D'Agostini:1994zf}, as implemented in the \textsc{RooUnfold} package~\cite{2011arXiv1105.1160A}, with a single iteration. The response matrix is obtained with \PYTHIAMBR ($\varepsilon = 0.08$). The cross section is measured only for $\Delta\eta^\mathrm{F} < 8.4$, so as to avoid regions where the trigger inefficiency and the unfolding uncertainty are large.

\subsection{Corrected results}\label{sec:correctedresults}

Figure~\ref{fig:corrected} shows the unfolded and fully corrected differential cross section $\rd\sigma / \rd\Delta\eta^\mathrm{F}$ for events with at least one particle with $\pt > 200$\MeV in the region of $\abs{\eta}<4.7$. As the statistical uncertainty is negligible, only the systematic uncertainty, discussed in Section~\ref{sec:syst}, is shown. The predictions from \PYTHIAMBR ($\varepsilon = 0.08$ and $0.104$), \PYTHIA8 tune 4C, and \PYTHIA6 tune Z2$^*$ are also given. The MC predictions show that in the pseudorapidity range covered by the measurement, $\abs{\eta}<4.7$, a large fraction of nondiffractive events can be suppressed by means of the $\Delta\eta^\mathrm{F}>3$ requirement.

\begin{figure*}[!thb]
\centering
\includegraphics[width=0.49\textwidth]{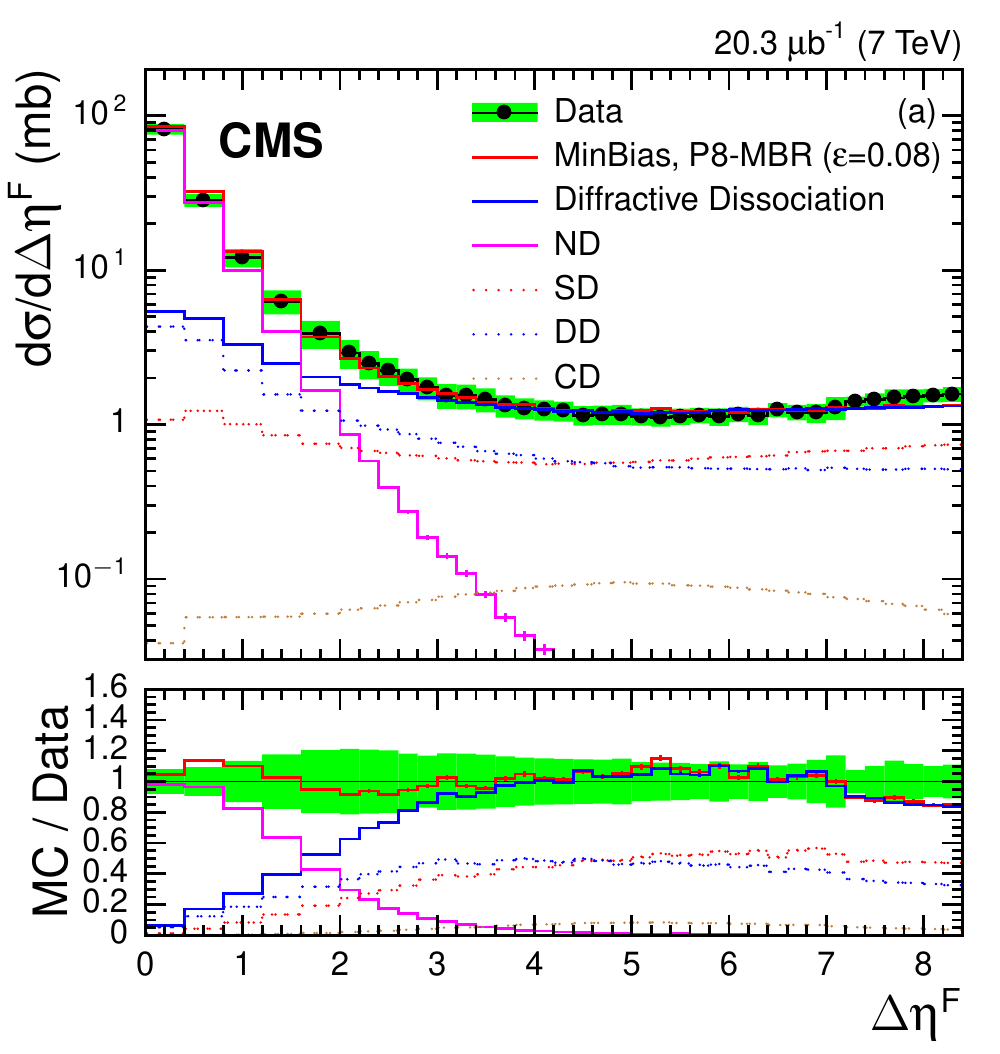}
\includegraphics[width=0.49\textwidth]{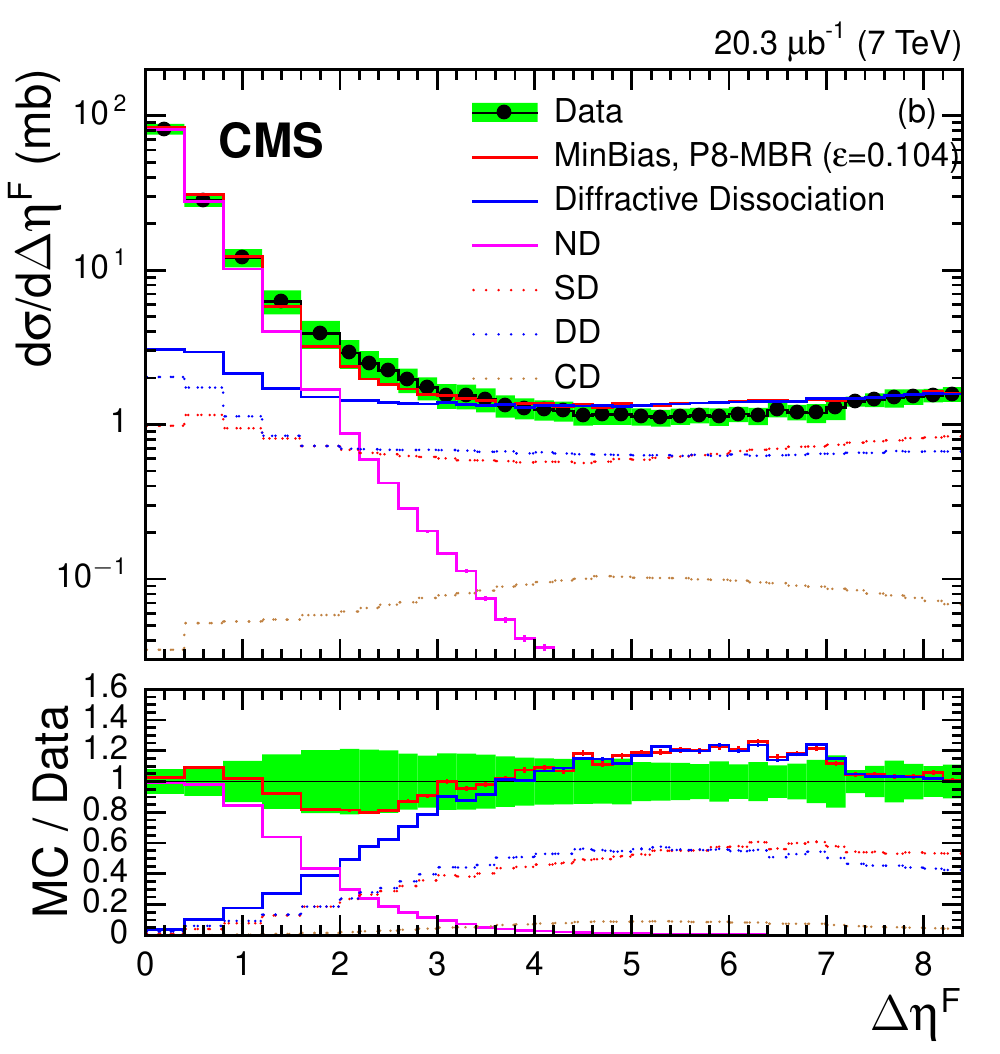}\\
\includegraphics[width=0.49\textwidth]{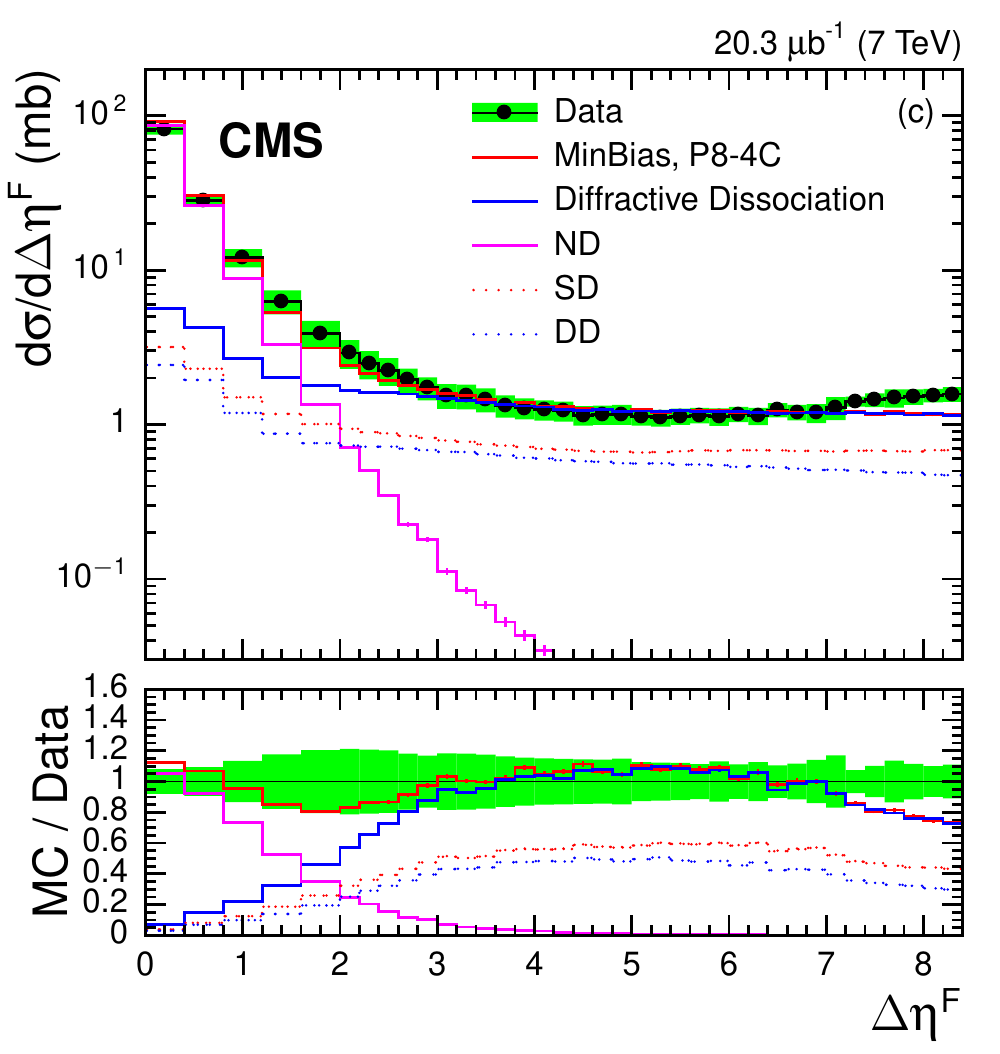}
\includegraphics[width=0.49\textwidth]{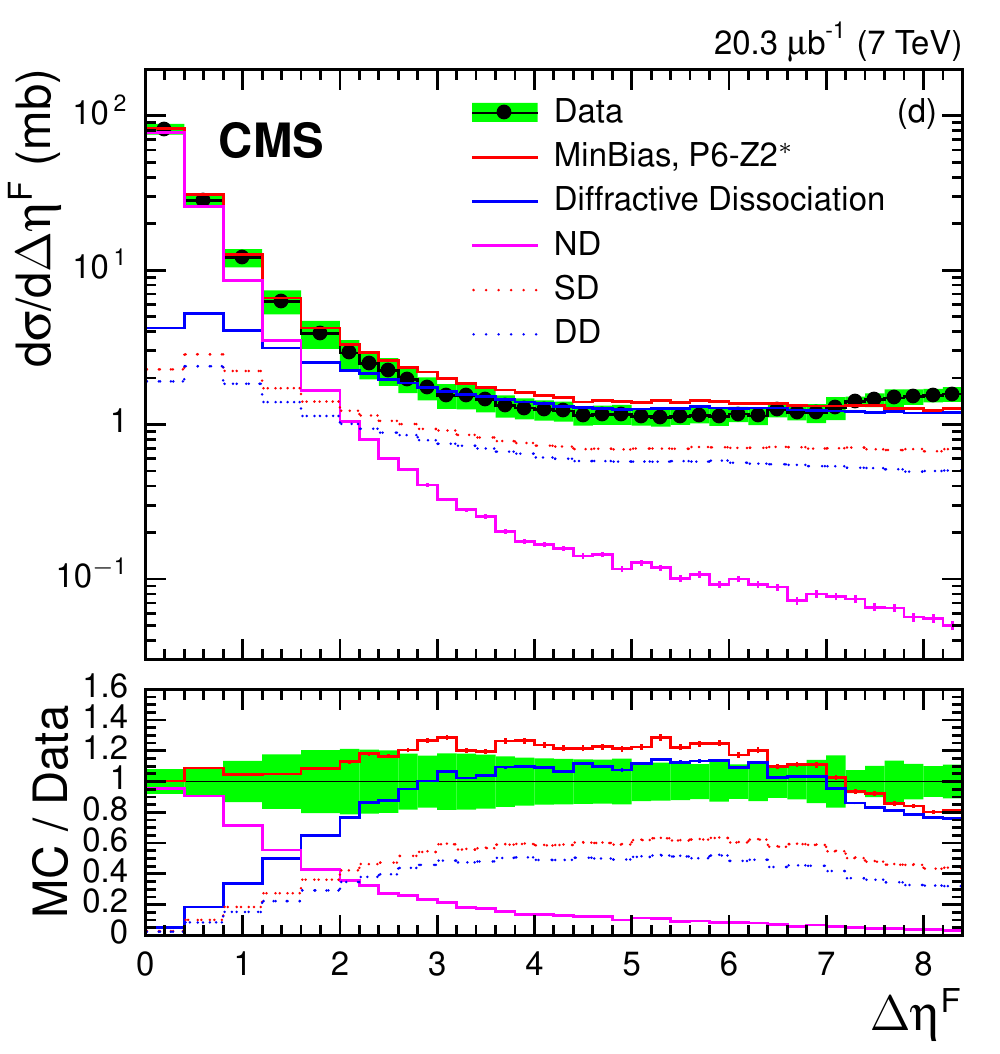}

\caption{Differential cross section $\rd\sigma / \rd\Delta\eta^\mathrm{F}$ for stable particles with $\pt>200$\MeV in the region $\abs{\eta} < 4.7$ compared to the corresponding predictions of (a) \PYTHIAMBR ($\varepsilon = 0.08$), (b) \PYTHIAMBR ($\varepsilon = 0.104$), (c) {\PYTHIA8} tune 4C, and (d) \PYTHIA6 tune Z2*. The band around the data points represents the total systematic uncertainty, which is discussed in Section~\ref{sec:syst}.}
\label{fig:corrected}
\end{figure*}

The present results are consistent with those from the ATLAS collaboration~\cite{Aad:2012pw}, as shown in Fig.~\ref{fig:finalatlas}. The ATLAS measurement uses all stable final-state particles with $\pt>200$\MeV over the region $\abs{\eta}<4.9$. According to the \PYTHIAMBR simulation, the difference in the $\eta$ coverage between the two experiments causes changes in the $\Delta\eta^\mathrm{F}$ distribution of up to 5\%.

\begin{figure}[thb]
\centering
\includegraphics[width=0.49\textwidth]{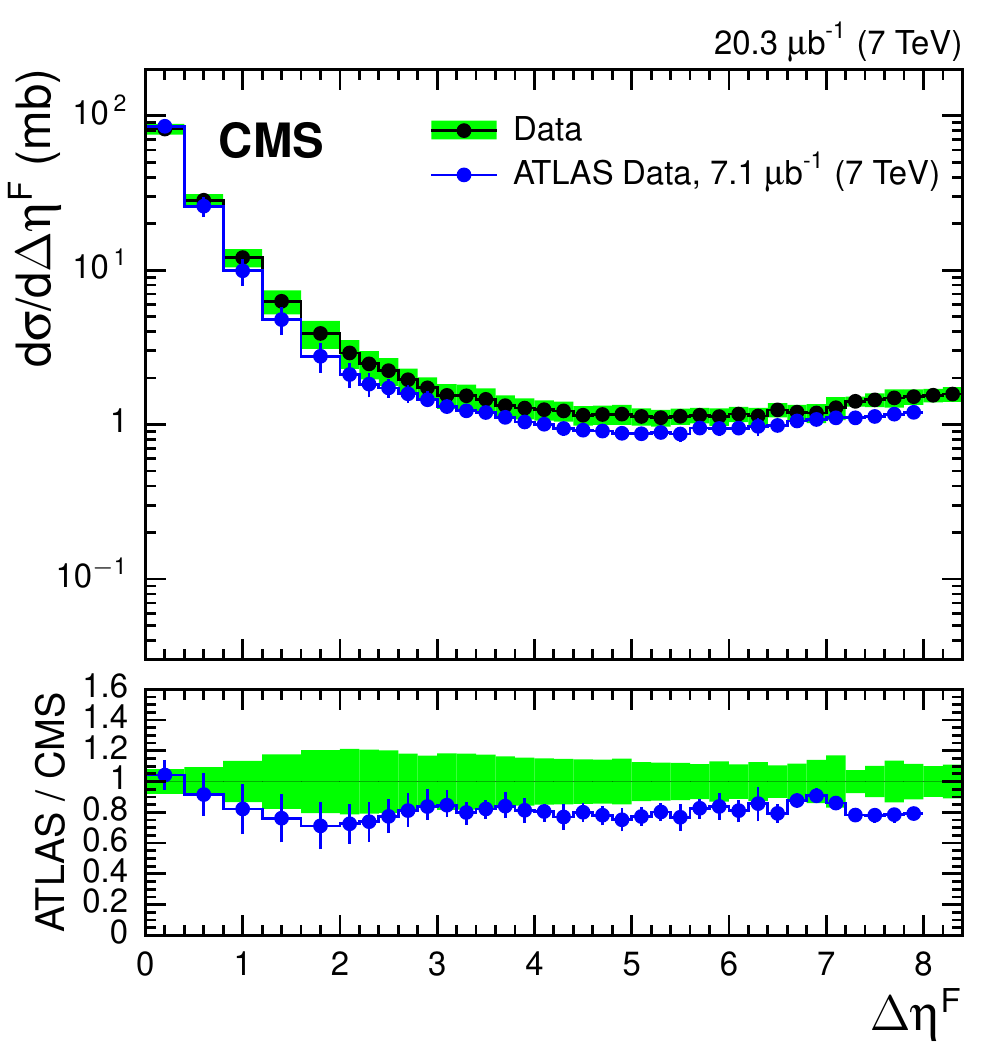}

\caption{Differential cross section $\rd\sigma / \rd\Delta\eta^\mathrm{F}$ for stable particles with $\pt>200$\MeV in the region $\abs{\eta} < 4.7$ compared to the ATLAS result~\cite{Aad:2012pw}: distributions (top) and ratio to the CMS measurement (bottom). The band represents the total systematic uncertainty in the CMS measurement, while the uncertainty in the ATLAS measurement is shown by the error bars. The stable-particle level definitions of the two measurements are not exactly identical: CMS measures the forward pseudorapidity gap size starting from $\eta = {\pm}4.7$, whereas the ATLAS limit is $\eta = {\pm}4.9$.}

\label{fig:finalatlas}
\end{figure}

\section{Systematic uncertainties}
\label{sec:syst}

Systematic uncertainties are obtained by varying the selection criteria and modifying the analysis. The following sources of systematic uncertainties are taken into account for the results presented in Sections~\ref{sec:sd}--\ref{sec:rapgapxsec}:

\begin{itemize}
\item HF energy scale: varied in the MC simulations by $\pm$10\%, to reflect the energy scale uncertainty estimated for data.

\item PF energy thresholds: raised by 10\%, based on dedicated studies of the detector noise.

\item Modeling of the diffractive interaction and the hadronization process: the hadronization parameters in the nominal \PYTHIAMBR MC sample are tuned to describe the multiplicity and $\pt$ spectra of diffractive systems in $\Pp\Pp$ and $\Pp \Pap$ collisions at $\sqrt{s} \le 1800$\GeV~\cite{DINOHAD}. The corresponding uncertainty is estimated by taking the difference (see Appendix B) between the results obtained with \PYTHIAMBR and those obtained with \PYTHIA8~4C (Sections~\ref{sec:sd}--\ref{sec:sdddsig}), or \PYTHIA8~4C and \PYTHIA6-Z2* (Section~\ref{sec:rapgapxsec}).

\item The uncertainty in the integrated luminosity measurement is $\pm$4\%~\cite{lumi1,lumi2}.

\end{itemize}

In addition, the following checks are carried out for the results shown in Sections~\ref{sec:sd}--\ref{sec:sdddsig}:

\begin{itemize}
\item CASTOR energy scale~\cite{CMS-DP-2013-035}: changed in the simulation by $\pm 20\%$, to reflect the estimated energy scale uncertainty for the data.

\item CASTOR energy threshold in each sector~\cite{Chatrchyan:2013gfi}: changed from the nominal 4$\sigma$ to 3.5$\sigma$ and 5$\sigma$, where $\sigma$ is the pedestal width.

\item CASTOR alignment uncertainty~\cite{CMS-DP-2014-014}: the simulated CASTOR position  in the plane transverse to the beamline is varied within the limits allowed by the condition that the MC description of the energy flow in CASTOR remains satisfactory. This corresponds to about 10 mm and 4 mm, for the left and right CASTOR sides, respectively.

\item Trigger efficiency uncertainty: estimated from a comparison of efficiency curves between data (measured by using a control sample for which no trigger requirements were applied) and MC.

\item Background subtraction: backgrounds from DD and ND events in the SD sample, and from ND and SD in the DD sample are estimated with \PYTHIAMBR (Figs.~\ref{fig:5} and \ref{fig:70}). The corresponding uncertainty is estimated by varying their relative contributions by $\pm 10\%$ (average normalization uncertainty of the model). The contribution from CD events in the SD and DD samples is negligible.

\end{itemize}

The total systematic uncertainty is obtained by summing all individual uncertainties in quadrature, separately for the positive and negative deviations from the nominal cross section values, leading to a total systematic uncertainty of up to 25\%. Table~\ref{tab:5} presents the summary of the systematic uncertainties for the measurement of the $\sigma^\mathrm{SDvis}$, $\sigma^\mathrm{DDvis}_\text{CASTOR}$, and $\sigma^\mathrm{DDvis}_\mathrm{CG}$ cross sections. The systematic uncertainties are significantly larger than the statistical ones, and the dominant sources are the HF energy scale and the modeling of diffraction and hadronization. For the $\sigma^\mathrm{DDvis}_\text{CASTOR}$ cross section, also the uncertainty related to the CASTOR alignment is significant.

\begin{table*}[htb]
\centering
\topcaption{Systematic uncertainties for the measurement of the $\sigma^\mathrm{SDvis}$, $\sigma^\mathrm{DDvis}_\text{CASTOR}$, and $\sigma^\mathrm{DDvis}_\mathrm{CG}$ cross sections; individual contributions, as well as total systematic and statistical uncertainties are shown.}
\label{tab:5}
\begin{scotch}{lccc}
 \multirow{2}{*}{Source} & \multicolumn{3}{c}{Uncertainty (\%)}\\ \cline{2-4}
\rule[1.2\baselineskip]{0pt}{2.8ex}  & $\sigma^\mathrm{SDvis}$& $\sigma^\mathrm{DDvis}_\text{CASTOR}$ & $\sigma^\mathrm{DDvis}_\mathrm{CG}$ \\
\hline
 HF energy scale & 10 & 1.6 & 23 \\
 PF thresholds & 0.8 & 0.4 & 6.9\\
Diff. and had. modeling& 10 & 4.3 & 0.4\\
Luminosity & 4 & 4 & 4\\
CASTOR energy scale& 0.5 & 0.9 & 0\\
CASTOR threshold & 0.9 & 2.8& 0 \\
CASTOR alignment & 2.6 & 8.3 & 0\\
Trigger & 0.6 & 0.6 & 0.7 \\
Background sub. & 4.3 & 0.4 & 1.3 \\
\hline
Total systematic & 16 & 11 & 25 \\
Statistical & 0.9 & 1.8 & 1.3 \\
\end{scotch}

\end{table*}

 \section{Summary}
\label{sec:sum}

Measurements of diffractive dissociation cross sections in $\Pp\Pp$ collisions at $\sqrt{s}=7$\TeV have been presented in kinematic regions defined by the masses $M_{\cPX}$ and $M_{\cPY}$ of the two final-state hadronic systems separated by the largest rapidity gap in the event.  Differential cross sections are measured as a function of $\xi_{\cPX}= M^2_{\cPX}/s$ in the region $-5.5<\log_{10}\xi_{\cPX}<-2.5$, for $\log_{10}M_{\cPY}<0.5$, dominated by single dissociation (SD), and $0.5<\log_{10}M_{\cPY}<1.1$, dominated by double dissociation (DD). The discrimination between the above two $M_\cPY$ regions is performed by means of the CASTOR forward calorimeter. The cross sections integrated over these regions are $\sigma_\text{no-CASTOR} = 2.99 \pm 0.02\stat {}_{-0.29}^{+0.32}\syst$\unit{mb} and $\sigma_\text{CASTOR} = 1.18 \pm 0.02\stat\pm 0.13\syst$\unit{mb}, respectively.

The inclusive $\Pp\Pp$ cross section is also measured as a function of the width of the central pseudorapidity gap, $\Delta\eta$, for $\Delta\eta>3$, $\log_{10}M_\cPX>1.1$, and $\log_{10}M_\cPY>1.1$ (dominated by DD contributions). The corresponding integrated cross section is $\sigma_\mathrm{CG}=0.58 \pm 0.01\stat {}_{-0.11}^{+0.13}\syst$\unit{mb}.

The SD and DD cross sections in the above three regions, extracted by means of the \PYTHIAMBR simulation, which provides a good description of the data, are $\sigma^\mathrm{SDvis} = 4.06 \pm 0.04 \stat {}_{-0.63}^{+0.69} \syst$\unit{mb} (accounting for both $\Pp\Pp\to \cPX\Pp$ and $\Pp\Pp\to \Pp \cPY$), $\sigma^\mathrm{DDvis}_\text{CASTOR} = 1.06 \pm 0.02 \stat \pm 0.12 \syst$\unit{mb}, and $\sigma^\mathrm{DDvis}_\mathrm{CG} = 0.56 \pm 0.01 \stat {}_{-0.13}^{+0.15} \syst$\unit{mb}, respectively.

Extrapolations of the SD and DD cross sections to the regions $\xi<0.05$ and $\Delta\eta>3$, performed with \PYTHIAMBR, yield $\sigma^\mathrm{SD}=8.84 \pm 0.08 \stat {}^{+1.49}_{-1.38} \syst {}^{+1.17}_{-0.37} \,(\mathrm{extrap})$\unit{mb} and $\sigma^\mathrm{DD} = 5.17 \pm 0.08 \stat {}^{+0.55}_{-0.57} \syst ^{+1.62}_{-0.51}\,\text{(extrap)}$\unit{mb}, respectively.

In addition, the inclusive differential cross section $\rd\sigma / \rd\Delta\eta^\mathrm{F}$ for events with a pseudorapidity gap adjacent to the edge of the detector is measured over 8.4 units of pseudorapidity.

These measurements are compared to results from other experiments as well as to phenomenological predictions. The data are consistent with the SD and DD cross sections weakly rising with energy, and provide new experimental constraints on the modeling of diffraction in hadronic interactions.

\begin{acknowledgments}
\hyphenation{Bundes-ministerium Forschungs-gemeinschaft Forschungs-zentren} We congratulate our colleagues in the CERN accelerator departments for the excellent performance of the LHC and thank the technical and administrative staffs at CERN and at other CMS institutes for their contributions to the success of the CMS effort. In addition, we gratefully acknowledge the computing centers and personnel of the Worldwide LHC Computing Grid for delivering so effectively the computing infrastructure essential to our analyses. Finally, we acknowledge the enduring support for the construction and operation of the LHC and the CMS detector provided by the following funding agencies: the Austrian Federal Ministry of Science, Research and Economy and the Austrian Science Fund; the Belgian Fonds de la Recherche Scientifique, and Fonds voor Wetenschappelijk Onderzoek; the Brazilian Funding Agencies (CNPq, CAPES, FAPERJ, and FAPESP); the Bulgarian Ministry of Education and Science; CERN; the Chinese Academy of Sciences, Ministry of Science and Technology, and National Natural Science Foundation of China; the Colombian Funding Agency (COLCIENCIAS); the Croatian Ministry of Science, Education and Sport, and the Croatian Science Foundation; the Research Promotion Foundation, Cyprus; the Ministry of Education and Research, Estonian Research Council via IUT23-4 and IUT23-6 and European Regional Development Fund, Estonia; the Academy of Finland, Finnish Ministry of Education and Culture, and Helsinki Institute of Physics; the Institut National de Physique Nucl\'eaire et de Physique des Particules~/~CNRS, and Commissariat \`a l'\'Energie Atomique et aux \'Energies Alternatives~/~CEA, France; the Bundesministerium f\"ur Bildung und Forschung, Deutsche Forschungsgemeinschaft, and Helmholtz-Gemeinschaft Deutscher Forschungszentren, Germany; the General Secretariat for Research and Technology, Greece; the National Scientific Research Foundation, and National Innovation Office, Hungary; the Department of Atomic Energy and the Department of Science and Technology, India; the Institute for Studies in Theoretical Physics and Mathematics, Iran; the Science Foundation, Ireland; the Istituto Nazionale di Fisica Nucleare, Italy; the Ministry of Science, ICT and Future Planning, and National Research Foundation (NRF), Republic of Korea; the Lithuanian Academy of Sciences; the Ministry of Education, and University of Malaya (Malaysia); the Mexican Funding Agencies (CINVESTAV, CONACYT, SEP, and UASLP-FAI); the Ministry of Business, Innovation and Employment, New Zealand; the Pakistan Atomic Energy Commission; the Ministry of Science and Higher Education and the National Science Centre, Poland; the Funda\c{c}\~ao para a Ci\^encia e a Tecnologia, Portugal; JINR, Dubna; the Ministry of Education and Science of the Russian Federation, the Federal Agency of Atomic Energy of the Russian Federation, Russian Academy of Sciences, and the Russian Foundation for Basic Research; the Ministry of Education, Science and Technological Development of Serbia; the Secretar\'{\i}a de Estado de Investigaci\'on, Desarrollo e Innovaci\'on and Programa Consolider-Ingenio 2010, Spain; the Swiss Funding Agencies (ETH Board, ETH Zurich, PSI, SNF, UniZH, Canton Zurich, and SER); the Ministry of Science and Technology, Taipei; the Thailand Center of Excellence in Physics, the Institute for the Promotion of Teaching Science and Technology of Thailand, Special Task Force for Activating Research and the National Science and Technology Development Agency of Thailand; the Scientific and Technical Research Council of Turkey, and Turkish Atomic Energy Authority; the National Academy of Sciences of Ukraine, and State Fund for Fundamental Researches, Ukraine; the Science and Technology Facilities Council, UK; the US Department of Energy, and the US National Science Foundation.

Individuals have received support from the Marie-Curie program and the European Research Council and EPLANET (European Union); the Leventis Foundation; the A. P. Sloan Foundation; the Alexander von Humboldt Foundation; the Belgian Federal Science Policy Office; the Fonds pour la Formation \`a la Recherche dans l'Industrie et dans l'Agriculture (FRIA-Belgium); the Agentschap voor Innovatie door Wetenschap en Technologie (IWT-Belgium); the Ministry of Education, Youth and Sports (MEYS) of the Czech Republic; the Council of Science and Industrial Research, India; the HOMING PLUS program of the Foundation for Polish Science, cofinanced from European Union, Regional Development Fund; the Compagnia di San Paolo (Torino); the Consorzio per la Fisica (Trieste); MIUR project 20108T4XTM (Italy); the Thalis and Aristeia programs cofinanced by EU-ESF and the Greek NSRF; and the National Priorities Research Program by Qatar National Research Fund.
\end{acknowledgments}
\bibliography{auto_generated}
\clearpage
\appendix\section{SD/DD extrapolation factors}

\begin{figure*}[b!]
\centering
\includegraphics[width=0.9\textwidth]{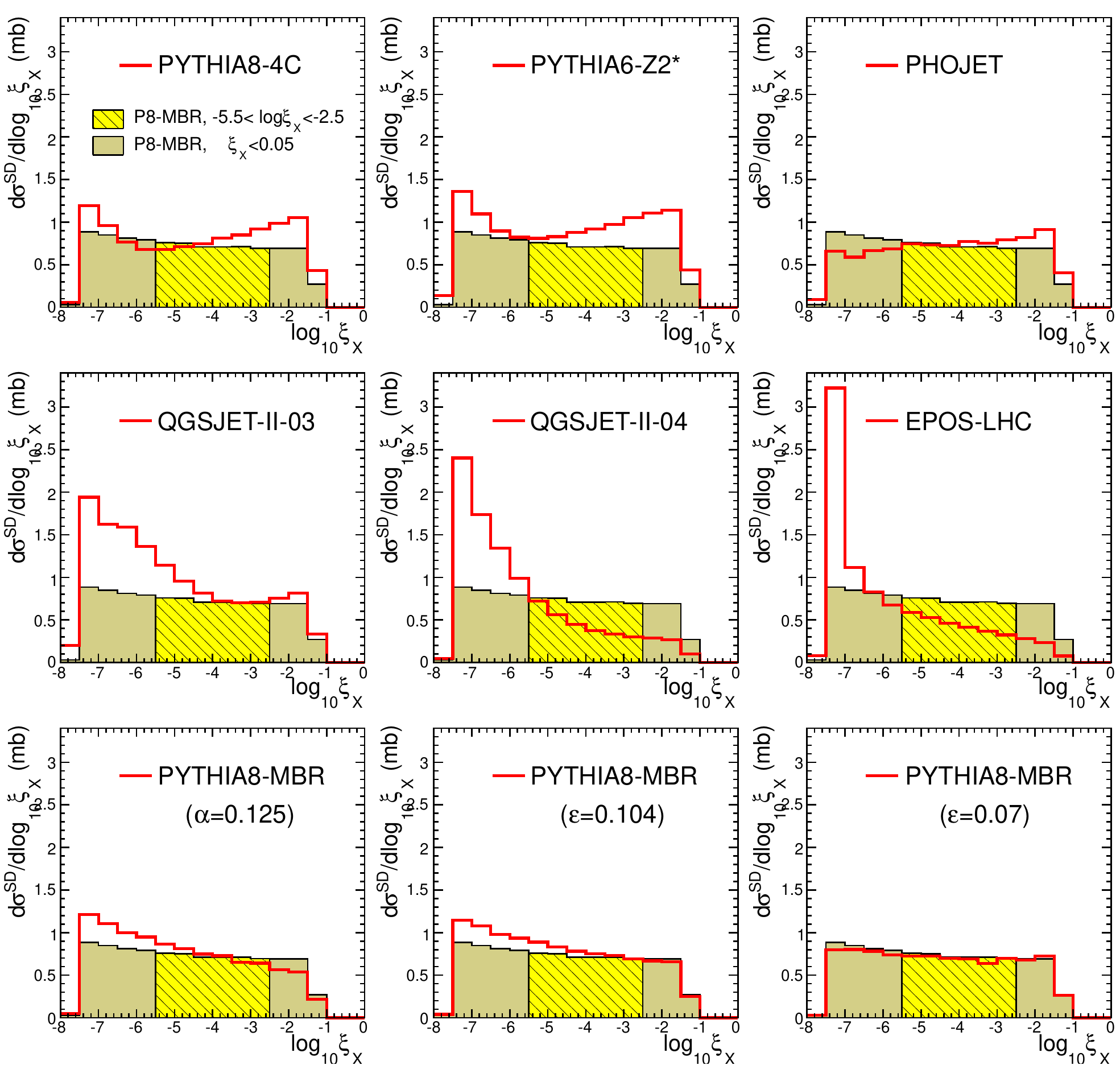}
\caption{Generator-level SD cross section as a function of $\xi_\cPX=M^2_\cPX/s$ for $\xi_\cPX<0.05$, shown for \PYTHIA8~4C, \PYTHIA6~Z2*, \PHOJET, \QGSJETII~03, \QGSJETII~04, \EPOS MC, and \PYTHIAMBR with the parameters of the Pomeron trajectory changed from the nominal values ($\alpha'= 0.25\GeV^{-2},~\varepsilon = 0.08$) to $\alpha' = 0.125\GeV^{-2}$, $\varepsilon=0.104$, and $\varepsilon=0.07$ (one parameter changed at a time). The nominal \PYTHIAMBR simulation is presented in each plot for the two regions of $\xi_\cPX$, $-5.5<\log_{10}\xi_\cPX<-2.5$ (dashed yellow) and $\xi_\cPX<0.05$ (solid khaki), used to extrapolate the measured SD cross section (from the dashed (yellow) to the solid (khaki) regions).}
\label{fig:extrap}

\end{figure*}

Figure~\ref{fig:extrap} shows the $\xi_\cPX=M^2_\cPX/s$ dependence of the SD cross section for the \PYTHIA8~4C, \PYTHIA6~Z2*, \PHOJET, \QGSJETII~03, \QGSJETII~04, and \EPOS simulations, compared to the nominal \PYTHIAMBR simulation used in this analysis for two regions of $\xi_\cPX$, $-5.5<\log_{10}\xi_\cPX<-2.5$ (dashed yellow) and $\xi_\cPX<0.05$ (solid khaki). In addition, the \PYTHIAMBR simulations with values of $\alpha'$ and $\varepsilon$ changed to $\alpha' = 0.125\GeV^{-2}$, $\varepsilon=0.104$, and $\varepsilon=0.07$ (one parameter changed at a time) are also included to provide a scale for their effect on the cross sections. Extrapolation factors, defined as the ratios of $\sigma^\mathrm{SD}(\xi_\cPX<0.05)$ to $\sigma^\mathrm{SD,vis}(-5.5<\log_{10}\xi_\cPX<-2.5)$, are presented for each of the above ten MC models in Table~\ref{tab:fsd}. For each model, two ratios are evaluated, one in which both cross sections (numerator and denominator of the extrapolation factor) are calculated by using the same generator ($f^\mathrm{SD}$), and another where the prediction of the $\sigma^\mathrm{SD,vis}(-5.5<\log_{10}\xi_\cPX<-2.5)$ is calculated by using the nominal \PYTHIAMBR with $\varepsilon = 0.08$ ($f^\mathrm{SD}_\mathrm{MBR}$). The numbers in brackets in Table~\ref{tab:fsd} show the relative change of the extrapolation factors with respect to the nominal \PYTHIAMBR simulation; the one related to $f^\mathrm{SD}$ is sensitive to the difference in the shape of the $\xi_\cPX$ (mass) distribution, while the one related to $f^\mathrm{SD}_\mathrm{MBR}$ is also sensitive to the the normalization of the SD cross section.

Table~\ref{tab:fdd} shows the extrapolation factors for the DD case, defined as the ratios of $\sigma^\mathrm{DD}(\Delta\eta>3)$ to $\sigma^\mathrm{DD,vis}$, again for two cases: one when both cross sections are calculated with the same MC generator ($f^\mathrm{DD}$), and the other when the predicted $\sigma^\mathrm{DD,vis}$ cross section is from \PYTHIAMBR with $\varepsilon = 0.08$ ($f^\mathrm{DD}_\mathrm{MBR}$).
 The relative change of the extrapolation factors with respect to the nominal \PYTHIAMBR simulation is shown in brackets; the one related to $f^\mathrm{DD}$ accounts for the difference in the shape of the $M_\cPX$ and $M_\cPY$ dependence, while the one related to $f^\mathrm{DD}_\mathrm{MBR}$ is sensitive to the difference in the mass dependence and the normalization of the DD cross section.

\begin{table}[t!hb]
\centering
\topcaption{Extrapolation factors $f^\mathrm{SD}=\sigma^\mathrm{SD}_\mathrm{i}(\xi<0.05)/\sigma^\mathrm{SD,vis}_\mathrm{i}$ and $f^\mathrm{SD}_\mathrm{MBR}=\sigma^\mathrm{SD}_\mathrm{i}(\xi<0.05)/\sigma^\mathrm{SD,vis}_\mathrm{MBR}$ from the visible to total SD ($\xi<0.05$) cross section for each MC model considered ($i=1-10$). The relative change with respect to the value obtained by \PYTHIAMBR with $\varepsilon = 0.08$ is shown in parenthesis.}
\label{tab:fsd}
\begin{scotch}{cccccc}
$i$ & MC model & $f^\mathrm{SD}$ & $f^\mathrm{SD}_\mathrm{MBR}$\\
\hline
1& \PYTHIAMBR ($\varepsilon=0.08$)& 2.18\,(1.00) & 2.18\,(1.00)\\
\hline
2& \PYTHIA8~4C & 2.32\,(1.06) & 2.51\,(1.15)\\
3& \PYTHIA6~Z2* & 2.29\,(1.06) & 2.89\,(1.34)\\
\hline
4& \PHOJET & 2.06\,(0.95) & 2.18\,(1.00)\\
5& \QGSJETII~03 & 2.72\,(1.25) & 3.19\,(1.46)\\
6& \QGSJETII~04 & 3.62\,(1.66) & 2.30\,(1.06)\\
7& \EPOS & 3.44\,(1.58) & 2.15\,(0.99)\\
\hline
8& \PYTHIAMBR ($\alpha'=0.125$) & 2.27\,(1.04) & 2.34\,(1.07)\\
9& \PYTHIAMBR ($\varepsilon=0.104$)& 2.23\,(1.03) & 2.42\,(1.11)\\
10& \PYTHIAMBR ($\varepsilon=0.07$)& 2.16\,(0.99) & 2.09\,(0.96)\\
\end{scotch}
\end{table}

\begin{table}[b!th]
\centering
\topcaption{Extrapolation factors $f^\mathrm{DD}=\sigma^\mathrm{DD}_\mathrm{i}(\Delta\eta>3)/\sigma^\mathrm{DD,vis}_\mathrm{i}$ and $f^\mathrm{DD}_\mathrm{MBR}=\sigma^\mathrm{DD}_\mathrm{i}(\Delta\eta>3)/\sigma^\mathrm{DD,vis}_\mathrm{MBR}$ from the visible to total DD ($\Delta\eta>3$) cross section for each MC model considered ($i=1-10$). The relative change with respect to the value obtained by \PYTHIAMBR with $\varepsilon = 0.08$ is shown in parenthesis.}
\label{tab:fdd}
\begin{scotch}{cccccc}
$i$ & MC model & \multicolumn{2}{c}{$f^\mathrm{DD}$} & \multicolumn{2}{c}{$f^\mathrm{DD}_\mathrm{MBR}$}\\
\hline
1& \PYTHIAMBR ($\varepsilon=0.08$) & 1.92~~(1.00) && 1.92~~(1.00)\\
\hline
2& \PYTHIA8~4C & 2.52\,(1.32) && 1.86\,(0.97)\\
3& \PYTHIA6~Z2* & 2.39\,(1.25) && 2.15\,(1.13)\\
\hline
4& \PHOJET & 1.80\,(0.94) && 0.60\,(0.31)\\
5& \QGSJETII~03 &\NA&& \NA&\\
6& \QGSJETII~04 & 2.04\,(1.07) && 0.94\,(0.49)\\
7& \EPOS & 4.73\,(2.47) && 1.93\,(1.01)\\
\hline
8& \PYTHIAMBR ($\alpha'=0.125$) & 1.97\,(1.03) && 2.32\,(1.21)\\
9& \PYTHIAMBR ($\varepsilon=0.104$) & 2.00\,(1.04) && 2.37\,(1.24)\\
10& \PYTHIAMBR ($\varepsilon=0.07$) & 1.88\,(0.98) && 1.73\,(0.90)\\
\end{scotch}
\end{table}

\section{MC hadronization models}

In this section, the hadronization models used to generate particle spectra in the simulations introduced in Section~\ref{sec:mc} are compared to a reference model~\cite{DINOHAD,ua1} based on data. The model correctly describes the charged-particle multiplicity and $\pt$ spectra of diffractive and inclusive proton-(anti)proton data for $\sqrt{s} \le 1800$\GeV by assuming that the Pomeron-proton collision produces a system of mass $M_\cPX$ that hadronizes as if it had been produced in a nondiffractive proton-proton collision at $\sqrt{s}=M_\cPX$. Figures~\ref{fig:hadn} and \ref{fig:hadpt} show the charged-particle multiplicity distributions and $\pt$ spectra for the SD process for three ranges of $M_\cPX$: $5.6<M_\cPX<10$\GeV, $32<M_\cPX<56$\GeV, and $178<M_\cPX<316$\GeV, for the \PYTHIAMBR, \PYTHIA8~4C, \PYTHIA6~Z2*, \PHOJET, \QGSJETII~03, \QGSJETII~04, and \EPOS simulations, compared to the reference model quoted above. The following conclusions can be drawn: \PYTHIA6~Z2*, \QGSJETII~03, \QGSJETII~04, and \EPOS predict smaller multiplicities and harder $\pt$ spectra than the model of Ref.~\cite{DINOHAD,ua1}; \PYTHIA8~4C agrees with the mean values of the multiplicity distributions, but predicts narrower widths, and harder $\pt$ spectra; \PHOJET predicts multiplicity distributions consistent with the reference model with harder $\pt$ spectra; and \PYTHIAMBR agrees with the reference model in both multiplicity and $\pt$ spectra. The latter is an expected result, as the hadronization parameters of the \PYTHIAMBR simulation have been tuned to follow the reference model.

\begin{figure*}[b!ht]
\centering
\includegraphics[width=0.8\textwidth]{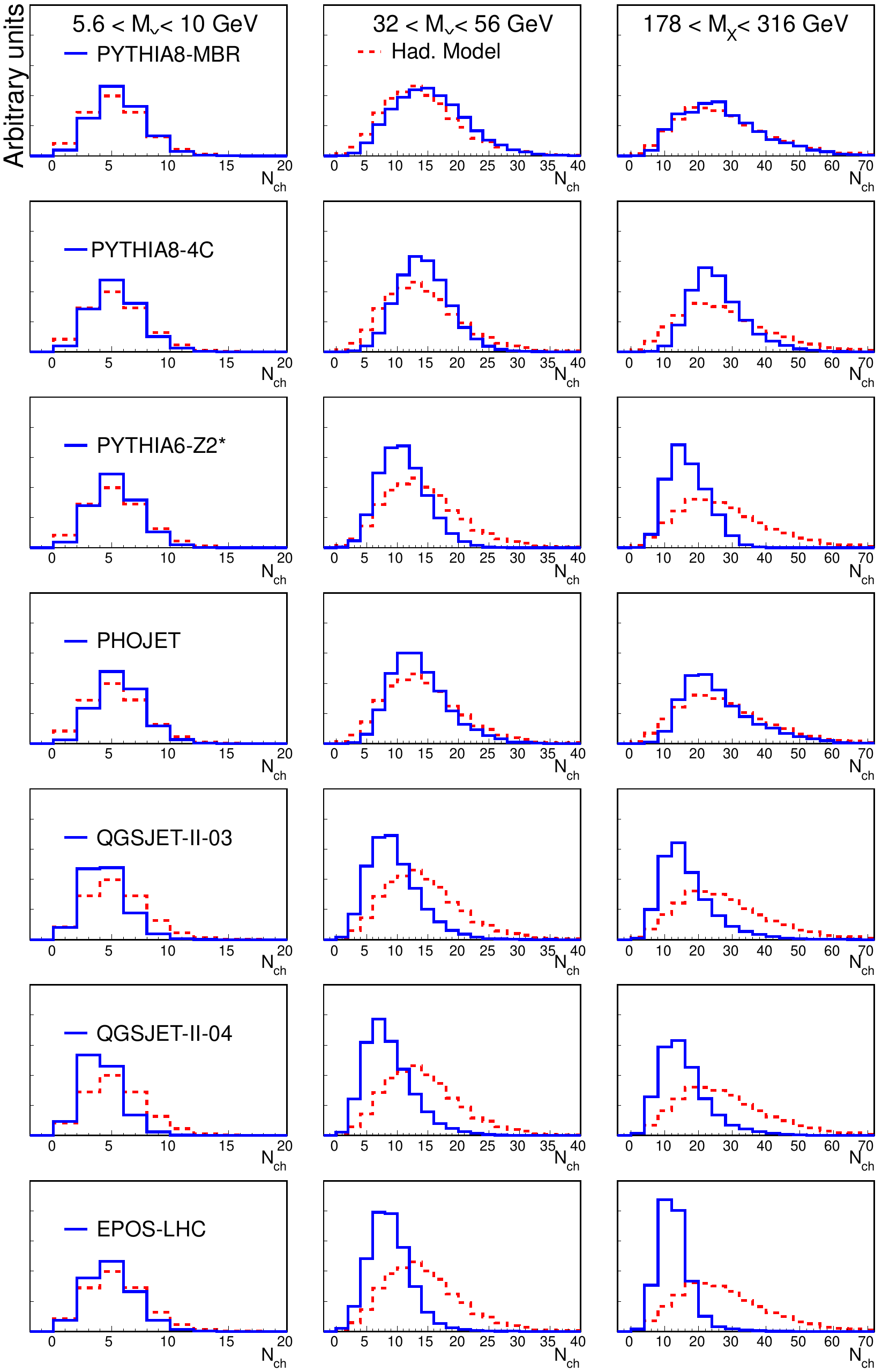}
\caption{Charged-particle multiplicity ($N_\mathrm{ch}$) distributions (area-normalized) in the \PYTHIAMBR, \PYTHIA8~4C, \PYTHIA6~Z2*, \PHOJET, \QGSJETII~03, \QGSJETII~04, and \EPOS MC simulations (rows) in three bins of $M_\cPX$ (columns) in SD collisions,
 compared to a reference hadronization model (dashed line), which describes the available data at $\sqrt{s} \le 1800$\GeV~\cite{DINOHAD,ua1}.}
\label{fig:hadn}
\end{figure*}

\begin{figure*}[thb]
\centering
\includegraphics[width=0.8\textwidth]{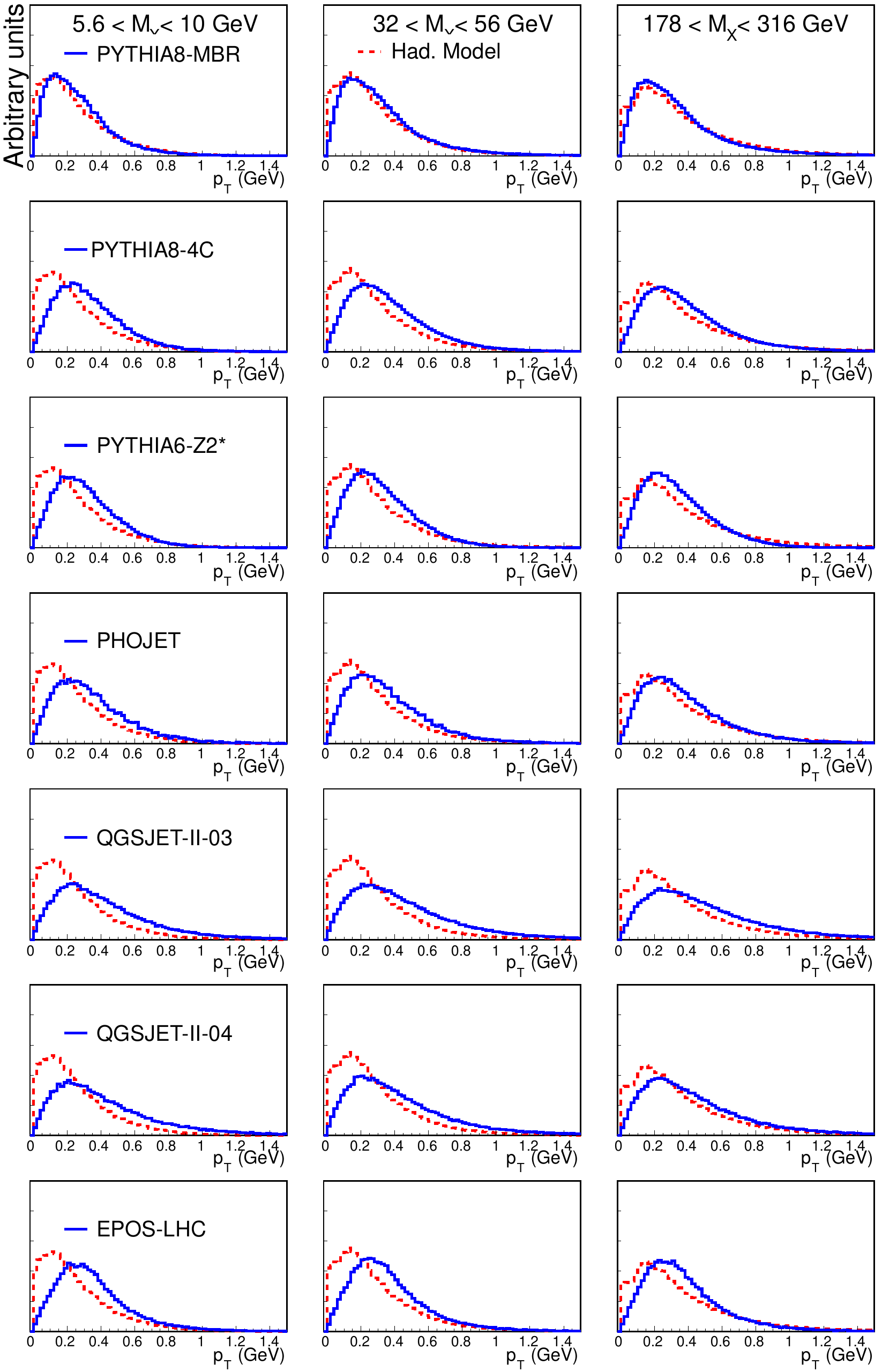}
\caption{Transverse-momentum ($\pt$) distributions (area-normalized) in the \PYTHIAMBR, \PYTHIA8~4C, \PYTHIA6~Z2*, \PHOJET, \QGSJETII~03, \QGSJETII~04, \EPOS MC simulations (rows) in three bins of $M_\cPX$ (columns) in SD collisions,
 compared to a reference hadronization model (dashed line), which describes the available data at $\sqrt{s} \le 1800$\GeV~\cite{DINOHAD,ua1}.}
\label{fig:hadpt}
\end{figure*}

\cleardoublepage \section{The CMS Collaboration \label{app:collab}}\begin{sloppypar}\hyphenpenalty=5000\widowpenalty=500\clubpenalty=5000\input{FSQ-12-005-authorlist.tex}\end{sloppypar}
\end{document}

%% file: FSQ-12-005-authorlist.tex
\textbf{Yerevan Physics Institute,  Yerevan,  Armenia}\\*[0pt]
V.~Khachatryan, A.M.~Sirunyan, A.~Tumasyan
\vskip\cmsinstskip
\textbf{Institut f\"{u}r Hochenergiephysik der OeAW,  Wien,  Austria}\\*[0pt]
W.~Adam, T.~Bergauer, M.~Dragicevic, J.~Er\"{o}, C.~Fabjan\cmsAuthorMark{1}, M.~Friedl, R.~Fr\"{u}hwirth\cmsAuthorMark{1}, V.M.~Ghete, C.~Hartl, N.~H\"{o}rmann, J.~Hrubec, M.~Jeitler\cmsAuthorMark{1}, W.~Kiesenhofer, V.~Kn\"{u}nz, M.~Krammer\cmsAuthorMark{1}, I.~Kr\"{a}tschmer, D.~Liko, I.~Mikulec, D.~Rabady\cmsAuthorMark{2}, B.~Rahbaran, H.~Rohringer, R.~Sch\"{o}fbeck, J.~Strauss, A.~Taurok, W.~Treberer-Treberspurg, W.~Waltenberger, C.-E.~Wulz\cmsAuthorMark{1}
% ...
\vskip\cmsinstskip
% ...
1:~~Also at Vienna University of Technology, Vienna, Austria\\
2:~~Also at CERN, European Organization for Nuclear Research, Geneva, Switzerland\\
% ...
